\documentclass[twocolumn, notitlepage, 10pt, aps, floatfix, showpacs, prb, citeautoscript, superscriptaddress, groupedaddress]{revtex4-1}
\usepackage{graphicx}
\usepackage{amsmath, amssymb}
\usepackage{bm}
\usepackage{bbm}
\usepackage{xcolor}
\usepackage[colorlinks, citecolor={blue!50!black}, urlcolor={blue!50!black}, linkcolor={red!50!black}]{hyperref}
\usepackage{bookmark}
\usepackage[version=4]{mhchem}
\usepackage{microtype}
\usepackage[load=physical,load=abbr, range-units=single]{siunitx}

\newcommand{\pmat}[1]{\begin{pmatrix}#1\end{pmatrix}}
\newcommand{\comment}[1]{}
\newcommand{\sgn}{\textrm{sgn}}

\begin{document}
\title{Topological superconductivity from magnetic impurities on monolayer \texorpdfstring{\ce{NbSe2}}{NbSe2}}

\author{Doru Sticlet}
\email{doru.sticlet@itim-cj.ro}
\affiliation{National Institute for Research and Development of Isotopic and Molecular Technologies, 67-103 Donat, 400293 Cluj-Napoca, Romania
}
\author{Cristian Morari}
\affiliation{National Institute for Research and Development of Isotopic and Molecular Technologies, 67-103 Donat, 400293 Cluj-Napoca, Romania
}

\begin{abstract}
Recent experimental studies have found that magnetic impurities deposited on superconducting monolayer \ce{NbSe2} generate coupled Yu-Shiba-Rusinov bound states.
Here we consider ferromagnetic chains of impurities which induce a Yu-Shiba-Rusinov band and harbor Majorana bound states at the chain edges.
We show that these topological phases are stabilized by strong Ising spin-orbit coupling in the monolayer and examine the conditions under which Majorana phases appear as a function of distance between impurities, impurity spin projection, orientation of chains on the surface of the monolayer, and strength of magnetic exchange energy between impurity and superconductor.
\end{abstract}

\maketitle

\section{Introduction}
Topological superconductivity in one dimension is characterized by the presence of zero-energy Majorana bound states (MBS) which encode nonlocally electronic degrees of freedom.~\cite{Kitaev2001}
This property, together with the non-Abelian statistics of their exchange, holds great promise to realize fault-tolerant topological quantum computation, where the coherence of qubits built from MBS is protected against local perturbations.~\cite{Kitaev2006, Nayak2008, Alicea2012, Leijnse2012, Beenakker2013, Sarma2015, Elliott2015, Sato2016, Aguado2017, Lutchyn2018}

The first experimental signatures of MBS were recorded in conductance measurements of semiconducting nanowires proximitized by $s$-wave superconductors~\cite{Mourik2012, Lutchyn2010, Oreg2010}.
Nevertheless, studies have proposed a wealth of additional platforms for the realization of one-dimensional $p$-wave topological superconductivity with MBS, including topological insulator-superconductor hybrids,~\cite{Fu2009a, Cook2011} cold atom quantum wires~\cite{Jiang2011}, proximitized half-metallic ferromagnets~\cite{Duckheim2011}, nanomagnets on superconductors,~\cite{Kjaergaard2012} etc.
The present paper focuses on proposals which generate MBS through the interplay between magnetic impurities and superconductivity.

\begin{figure}[t]
\includegraphics[width=0.9\columnwidth]{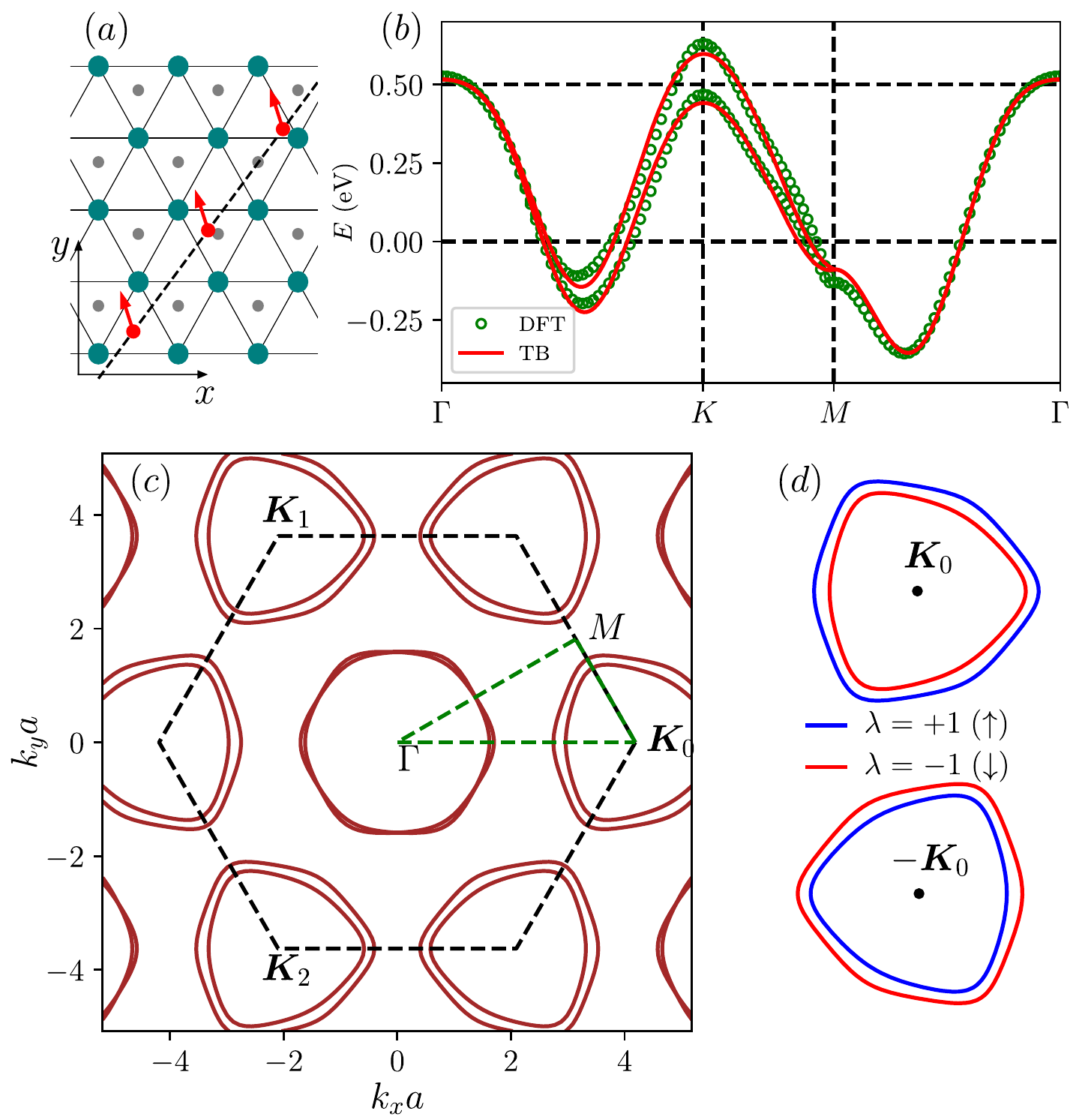}
\caption{(a) Monolayer \ce{NbSe2} viewed from the top.
The low-energy physics is described by an effective model for the triangular lattice formed by \ce{Nb} atoms (teal) (with \ce{Se} atoms in gray).
The dashed line marks the direction of a periodic chain of magnetic impurities (red dots) with an in-plane projection of spin (arrows).
(b) Energy dispersion along the high-symmetry lines in the Brillouin zone from \textit{ab initio} calculations (green circles, DFT) and the fit of the band structure with the one-band tight-binding model (red, TB).
(c) Fermi surface of metallic \ce{NbSe2} obtained from the tight-binding model, showing pockets at the $\Gamma$ point and at the corners $\pm \bm K_n$ of the Brillouin zone, with $n\in \{0, 1, 2\}$.
(d) Zoom at the $K$ valleys shows the split of the energy sheets.
Strong Ising spin-orbit coupling allows to index the Fermi sheets at the $K$ valleys with a helicity index $\lambda=\pm$.
The opposite valley changes the ordering sheets with respect to $\lambda$, under time-reversal symmetry.
}
\label{fig:fig1}
\end{figure}

The magnetic impurities break time-reversal and spin-rotation symmetries in $s$-wave superconductors and generate localized in-gap bound states, called Yu-Shiba-Rusinov (YSR) states~\cite{Yu1965, Shiba1968, Rusinov1969, Balatsky2006}, first detected experimentally for \ce{Mn} and \ce{Gd} adatoms on superconducting \ce{Nb}.~\cite{Yazdani1997}
When the impurities are close enough, the YSR states hybridize and form a band inside the superconducting gap.
Effective superconducting pairing may be induced in the band to generate stable topological phases.
Two theoretical avenues were envisaged to achieve this.
First, the superconductor substrate has no spin-orbit coupling, but helical ordering of impurity spins, opens a gap in the YSR band and generates effective topological $p$-wave superconductivity~\cite{Choy2011, Martin2012, Braunecker2013, Pientka2013, Klinovaja2013, Nadj-Perge2013, Poeyhoenen2014, Kim2014}.
Alternatively, the superconductors have a strong spin-orbit coupling, which allows ferro- and antiferromagnetic chains of impurities to exhibit MBS at the edges~\cite{Heimes2014, Heimes2015, Brydon2015, Kim2015, Kaladzhyan2016}.
In this second scenario, the inversion symmetry is broken in the superconductor and the order parameter becomes a mixture of singlet and triplet pairing,~\cite{Bauer2012, Yip2014, Samokhin2015, Smidman2017} with the latter essential to generate effective $p$-wave superconductivity in the chains.
Since the helical configuration of impurity spins proved harder to stabilize in experiments, the first detection of MBS from YSR states has followed the second avenue by employing chains of \ce{Fe} atoms deposited on superconducting \ce{Pb}~\cite{Nadj-Perge2014, Ruby2015, Pawlak2016, Feldman2016, Jeon2017}.
One experimental advantage of such setups is that scanning tunneling microscopy (STM) allowed direct imaging of the zero-energy local density of states characteristic for MBS, and, in principle, could be used to carry out braiding of MBS.~\cite{Li2016}
The spin-polarized STM is also used as a tool to detect MBS,~\cite{Jeon2017, Jaeck2019} since MBS wave functions have spin polarization sensitive to the impurity magnetization.~\cite{Sticlet2012, He2014, Kotetes2015, Chirla2016, Devillard2017, Li2018}

The search of topological superconducting phases supporting MBS has led lately to a surge in experimental investigations on YSR states~\cite{Heinrich2018}.
In particular, our theoretical study is prompted by the recent experimental breakthroughs in detecting YSR states in superconducting \ce{NbSe2}.~\cite{Menard2015, Kezilebieke2018}

The bulk \ce{NbSe2} is a layered transition metal dichalcogenide, with layers connected by weak van der Waals forces.
Each layer is composed of three covalently bonded planar triangular lattices, two of \ce{Se} atoms, with one of \ce{Nb} atoms intercalated between them, such that the unit cell has a trigonal prismatic structure.
The bulk \ce{NbSe2} is known to display a $3\times 3$ charge density wave ordering bellow $\SI{33}{K}$ and to become a superconductor bellow $\SI{7.2}{K}$.
Recent technical advances have allowed to separate out monolayers of \ce{NbSe2} from the bulk.~\cite{Novoselov2005, Cao2015}
It was shown that the critical temperature to the superconducting phase for monolayers is suppressed to approximately $1$~\cite{Wang2017} or $\SI{2}{K}$,~\cite{Cao2015, Ugeda2016}, while the critical temperature to realize charge density wave ordering increases to $\SI{145}{K}$ due to enhanced electron-phonon coupling.~\cite{Xi2015, Ugeda2016}
Since the monolayer has a thickness on the order of $\SI{1.2}{nm}$, it may be considered quasi-two-dimensional, placing \ce{NbSe2} in a select class of two-dimensional superconductors.
Moreover, superconductivity in monolayer \ce{NbSe2} was shown to have surprising features, most striking being its robustness to high in-plane magnetic fields.~\cite{Xi2016, Xing2017, Sohn2018}
This is due to the strong Ising spin-orbit coupling, which favors out-of-plane alignment of spins in opposite valleys.
This results in Cooper pairs with pinned out-of-plane spins, which, together with suppressed orbital response in the two-dimensional limit, leads to an enhanced in-plane critical magnetic field, factors of magnitude above the Pauli paramagnetic limit.~\cite{Xi2016, Xing2017, Sohn2018}
These experimental findings are not unique to \ce{NbSe2} and are shared by the group-VIB TMDs, with most notable representative being monolayer \ce{MoS2}.~\cite{Xiao2012, Taniguchi2012, Yuan2014, Lu2015, Shi2015, Saito2015, Costanzo2016}
While \ce{NbSe2} is metallic in its normal state, the latter are usually semiconductors which are gated into the conduction band before becoming superconductors.

The presence of a strong spin-orbit coupling in the superconductor was shown above to be an essential ingredient for a class of proposals to realize topological superconductivity.
This has motivated the recent experimental investigations of YSR states in \ce{NbSe2}.
In-gap bound states were generated either from Fe adatoms~\cite{Menard2015} or magnetic molecules (cobalt phthalocyianine)~\cite{Kezilebieke2018}.
There are two main reasons why this findings are of interest to us.
First, \ce{NbSe2} is a quasi-two-dimensional superconductor, which leads to a slower decay of YSR wave function in comparison with the three-dimensional case ($1/\sqrt r$ versus $1/r$, for distance smaller than the superconducting coherence length $\xi_0$).~\cite{Menard2015}
This allows to couple the localized bound states easier as shown in Ref.~\onlinecite{Kezilebieke2018}, where YSR dimers were first reported.
This advances pave the way towards realization of MBS from coupled YSR states in \ce{NbSe2}.
The second reason has to do with the exotic Ising spin-orbit coupling in \ce{NbSe2}.
Due to broken in-plane reflection symmetry in the quasi-two-dimensional plane, but conserved out-of-plane mirror symmetry, the spin-orbit coupling favors a spin alignment normal to the monolayer for electrons at the Fermi surface in $K$ valleys, which leads to robust superconductivity to in-plane magnetic fields. 
This is different from most cases studied before~\cite{Heimes2014, Heimes2015, Brydon2015, Kim2015, Kaladzhyan2016} where spin-orbit coupling is Rashba type, which favors in-plane orientation of spins.
For the latter, the depairing effect of in-plane magnetic fields is large, rendering problematic the manipulation of MBS with magnetic fields~\cite{Li2016}.

The theoretical studies on YSR states in superconducting \ce{NbSe2} have for now limited themselves to consider only a few magnetic impurities.
Besides the theoretical work accompanying Refs.~\onlinecite{Menard2015, Kezilebieke2018}, the interest in the effect of magnetic impurities on \ce{NbSe2} goes back to early 2000~\cite{Flatte2000}, where a model without spin-orbit coupling was studied.
More recently, the local density of states due to one to three impurities was investigated within the self-consistent formalism~\cite{Ptok2017}. Also the crossover from YSR states to Kondo impurity states came under focus~\cite{Kezilebieke2019}, as well as the interplay between charge density wave ordering and YSR states~\cite{Liebhaber2019}.
Instead, our study is greatly indebted and extends the recent theoretical works~\cite{Zhang2016, Sharma2016} on monolayer superconducting TMDs like \ce{MoS2}, which shares with \ce{NbSe2} the same $D_{3h}$ point-group symmetry.
These works show that topological superconductivity is stabilized by ferromagnetic chains of impurities in superconductors with Ising spin-orbit coupling.
The Fermi surface for \ce{NbSe2} is however quite different from gated \ce{MoS2} considered before,~\cite{Zhang2016, Sharma2016} with pockets not only at $K$, but also at $\Gamma$ valley [see Figs.~\ref{fig:fig1}(b) and~\ref{fig:fig1}(c)].
Moreover, the Fermi surface at $K$ valleys in \ce{NbSe2} are spin split, rendering the problem complicated as more scattering channels are allowed.
We show however that stable MBS are still possible, and provide a map of topological phases as a function of chain orientation on the two-dimensional superconducting substrate, the strength of magnetic exchange energy, and the distance between impurities.

The structure of the article is as follows.
Section~\ref{sec:model} introduces two effective models to describe superconductivity in monolayer \ce{NbSe2}.
The first model is an effective tight-binding Hamiltonian for the conduction band.
The model is mainly used in numerical simulations, and serves to derive the second, analytical model, which provides a better insight into generic features of topological superconductivity in \ce{NbSe2}.
Section~\ref{sec:ysr} investigates the addition of a periodic ferromagnetic chain of impurities in \ce{NbSe2} and obtains an effective model for the chain in the limit of dilute impurities, energetically close to the middle of the superconducting gap (the deep-impurity limit).
In the following, we obtain the topological phase diagram as a function of the distance between impurities and the orientation of impurity chains on the monolayer.
Section~\ref{sec:num_res} displays topological phases with MBS using numerical methods, confirming the previous analytical calculations.
The numerical simulations are done in the deep-impurity limit, and, beyond that, by using exact diagonalization of finite two-dimensional lattices.
The last concluding Sec.~\ref{sec:conc} discusses our results and envisages possible extensions.
Several appendices detail specific points in the paper regarding \textit{ab initio} simulations, tight-binding models, indirect exchange interactions between impurity spins, additional numerical results, characteristic lengths from our calculations, etc.

\section{Model for superconducting monolayer \texorpdfstring{\ce{NbSe2}}{NbSe2}}
\label{sec:model}
\subsection{Tight-binding model}
The band energy dispersion of \ce{NbSe2} from experiments\cite{Xi2016, Bawden2016, Xing2017} and \textit{ab initio} simulations (Appendix~\ref{app:abinit}) shows a single spin-split band crossing the Fermi level [Fig.~\ref{fig:fig1}(b)].
Since we aim to describe physics at the Fermi surface, at energy scales on the order of the superconducting gap $\Delta=\SI{1}{meV}$, far smaller than the bandwidth, it is necessary to devise a simple Hamiltonian to model electron dynamics in this band.
From orbital-resolved density of states it follows that the largest contribution to this low-energy band comes from $d$ orbitals on the \ce{Nb} atoms.
Therefore in a first approximation the effect of \ce{Se} atoms is incorporated in an effective Hamiltonian for the triangular sublattice of \ce{Nb} atoms~\cite{Liu2013, He2018}.

The Hamiltonian has to meet the symmetry requirements of the underlying triangular lattice [see Fig.~\ref{fig:fig1}(a)].
There is an in-plane threefold rotation symmetry represented by the operator $C_3=\exp(-\frac{2\pi i}{3}\sigma_z)$.
There is an out-of-plane mirror symmetry represented by $M_z=-i\sigma_z$.
There are three in-plane reflection symmetry axes, the one about the $y$ axis, and rotations of it under the threefold symmetry.
Characteristic for the TMDs and responsible for the Ising spin-orbit coupling is breaking three in-plane reflection symmetries (about $x$ axis and its threefold rotations).
Breaking the in-plane reflection symmetry generates a spin-orbit coupling characterized by an out-of-plane spin orientation.

The Bogoliubov-de Gennes (BdG) Hamiltonian modeling superconductivity in the band crossing the Fermi level reads
\begin{equation}
\mathcal H_0 = \frac{1}{2}\sum_{\bm k} C_{\bm k}^\dag H_0(\bm k) C_{\bm k},
\end{equation}
with the creation operators $C^\dag_{\bm k}=(c^\dag_{\bm k\uparrow}, c^\dag_{\bm k\downarrow}, c_{\bm k\downarrow}, -c_{\bm k\uparrow})$ and Bloch Hamiltonian:
\begin{equation}\label{bloch}
H_{0}(\bm k)=\xi_k\tau_z+\Lambda_{\bm k}\tau_z\sigma_z+\Delta\tau_x,
\end{equation}
with the Pauli matrices in spin space denoted by $\bm\sigma$, and $\bm \tau$, in particle-hole space.
The kinetic term $\xi_k$ follows from a Slater-Koster~\cite{Slater1954} expansion across the Brillouin zone.
Variations of this term up to different orders in number of nearest-neighbor hoppings have been used before in describing triangular lattices for TMDs.~\cite{Smith1985, Rossnagel2005, Inosov2008, Menard2015}
Here we have used up to fifth nearest-neighbor hopping terms (see Appendix~\ref{app:tb}).

In addition, a spin-orbit coupling term $\Lambda_{\bm k}$ is added to the model due to broken in-plane reflection symmetries.
This so-called Ising spin-orbit term favors an out-of-plane orientation of spins ($\sigma_z$) and it is responsible for the spin-splitting of the conduction band [Fig.~\ref{fig:fig1}(b)].
Since $\Lambda_{\bm k}$ is odd in momentum normal to the broken symmetry axes, it vanishes on the $\Gamma-M$ lines in the Brillouin zone (Appendix~\ref{app:tb}).
We will show that this leads to gapless superconducting phases when chains of magnetic impurities are oriented  along the unbroken symmetry axes, identical to the theoretical prediction in \ce{MoS2} TMDs~\cite{Zhang2016}.

Details about the structure of the Bloch Hamiltonian and the fitting parameters to the \textit{ab initio} data are relegated to Appendix~\ref{app:tb}, together with a derivation of the spin-orbit term $\Lambda_{\bm k}$, in Appendix~\ref{app:soc}.
Further comparison with a three-band model for metallic \ce{NbSe2} is contained in Appendix~\ref{app:3bnd}.

\subsection{Analytical model in the parabolic approximation}
In order to study the superconducting properties of \ce{NbSe2}, we derive an effective model for \ce{NbSe2} in its normal metallic phase.
Since the Fermi surface of \ce{NbSe2} shows electron pockets at $K$ and $\Gamma$ points, a valuable strategy is to obtain effective Hamiltonians around those points through a series expansion of the tight-binding Hamiltonian.
In order to make analytical progress, the ensuing Hamiltonians are further mapped to effective Hamiltonians with a parabolic energy dispersion. 

We start by showing the first terms at order $q^3$ near the high-symmetry points in the Brillouin zone, where $\bm q$ is momentum measured from either $\pm\bm K_n$ or $\Gamma$.
At the two $K$ valleys, the Hamiltonian reads
\begin{eqnarray}
H^{\eta K}(\bm q) &=& \bigg(\frac{\hbar^2 q^2}{2m_K}+\eta w q^3\cos(3\chi)-\mu_K\bigg)\\
&& + [\eta\kappa_0 + \eta\kappa_2 q^2 + \kappa_3 q^3\cos(3\chi)]\sigma_z + \mathcal O(q^4)
\notag,
\end{eqnarray}
where $q=|\bm q|$, the effective mass at $K$ valleys is $m_K$, with chemical potential $\mu_K$.
The valley index is denoted throughout by $\eta=\pm$, and the angle between $\bm q$ and the $x$ axis, by $\chi$.
The cubic term parametrized by $w$ represents the warping of the Fermi surface.
The spin-orbit coupling amplitude is parametrized by the set of $\kappa_i$, and it is consistent with low-energy continuum models developed for triangular lattices with an extremum at $K$ valleys as for \ce{MoS2, MoSe2,} or \ce{MoTe2}.~\cite{Wang2014, Wang2018}

For small momenta $\bm q$ near the $\Gamma$ point, the Hamiltonian reads
\begin{equation}
H^\Gamma (\bm q) = \frac{\hbar^2q^2}{2m_\Gamma} - \mu_\Gamma + \gamma_3 q^3\cos(3\chi)\sigma_z
+\mathcal O(q^4),
\end{equation}
recovering the continuum Hamiltonians derived in Ref.~\onlinecite{He2018}.

In the parabolic approximation, we restrict ourselves only at quadratic terms in momentum.
Therefore the \ce{NbSe2} valleys have circular Fermi surfaces and important quantities to our analysis such as Fermi momentum and velocity simplify to scalars.
In order to keep close to the experimental reality, we derive the effective Fermi momentum and velocity at each valley from the full tight-binding model with warped Fermi surfaces.
More precisely, the effective $v_F$ and $k_F$ are obtained form the angular average at the Fermi surface $\langle * \rangle=\frac{1}{2\pi}\int \!*\, d\chi$ (see Appendix~\ref{app:FSparams}).

The retarded Green's function for the superconductor reads:
\begin{equation}\label{tot_green}
G_{ij}(E) = \frac{1}{\Omega}\sum_{\bm k} e^{i\bm k\cdot \bm r_{ij}}
\frac{1}{E-H_0(\bm k)+i0^+},
\end{equation}
with $\bm r_{ij}=\bm r_i-\bm r_j$ and system size $\Omega$.
The sum carries over the total number of momentum states $\bm k$ in the first Brillouin zone.
For energies $E\sim\Delta\sim \SI{1}{meV}$, much smaller than the normal metal bandwidth ($\sim\SI{1}{eV}$), only the contribution from momenta at the Fermi surface is important.
Therefore the Green's function may be decomposed in a sum over $K$ and $\Gamma$ pockets:
\begin{equation}\label{decomp}
G_{ij}(E) \simeq G_{ij}^K(E) + G_{ij}^\Gamma(E),
\end{equation}
with $G_{ij}^K(E)$, the Green's function for electrons at $\pm\bm K_n$ valleys, and $G_{ij}^\Gamma (E)$, the Green's function at the $\Gamma$ valley.

The Green's function for the six $K$ valleys is modeled as:
\begin{equation}\label{GK}
G_{ij}^K(E) = \frac{1}{3\Omega}\sum_{n=0}^2\sum_{\eta,\lambda=\pm}\sum_{\bm q}
e^{i(\eta\bm K_n+\bm q)\cdot\bm r_{ij}} G^{\eta\lambda}(E,\bm q),
\end{equation}
where $\bm q$ is a small momentum near the $\pm\bm K_n$ valley.
The factor $1/3$ comes from an average over the three equivalent valleys for a fixed valley index $\eta$.
We have used the spectral decomposition of the Green's function in the helicity subspace $\lambda$~\cite{Gorkov2001} to obtain:
\begin{equation}\label{eq:gel}
G^{\eta\lambda}(E,\bm q)=\frac{1}{E-\xi_{q,\eta\lambda}\tau_z-\Delta\tau_x}
\frac{1+\lambda\sigma_z}{2}.
\end{equation}
The product is understood here as a Kronecker product between particle-hole space and spin space, with $1$ being interpreted when necessary as identity in the respective space.
The normal dispersion $\xi_{q,\eta\lambda}$ at the $\pm\bm K_n$ valleys is indexed by valley $\eta$ and helicity $\lambda$, which models polarized electrons with spin up ($\lambda=+$) and down ($\lambda=-$).
From now on, we will denote for notational simplicity the sums over valley and helicity as $\sum_{n\eta\lambda}$. 
The normal dispersion is approximated using the leading term $\kappa_0$ in spin-orbit coupling, $\xi_{q,\eta\lambda} = \hbar^2 q^2/2m_K-\mu_K+\eta\lambda \kappa_0$.
The second term ($\lambda\sigma_z$) in Eq.~\eqref{eq:gel} is due to spin-orbit coupling and generates spin-triplet superconducting correlations, necessary to engineer topological superconductivity.~\cite{Brydon2015}

At the $\Gamma$ pocket, there is no contribution from spin-orbit coupling to $\mathcal O(q^2)$.
Only the cubic term contributes in creating spin-triplet superconducting pairing and helps stabilizing topological nontrivial phases.
Nevertheless, the physics is dominated by the contribution from $K$ valleys, since the spin-orbit coupling at $\pm\bm K_n$ is by an order of magnitude larger than at $\Gamma$.~\footnote{from the tight-binding model, $\kappa_0\simeq \SI{0.078}{eV}$ and $\gamma_3/a^3\simeq \SI{0.0027}{eV}$, with $a=\SI{0.344}{nm}$ the distance between \ce{Nb} atoms.}
Therefore we neglect in the following the cubic contribution from the $\Gamma$ valley.
Under this approximation, the Green's function at the $\Gamma$ point reads:
\begin{equation}\label{GG}
G_{ij}^\Gamma = \frac{1}{\Omega}\sum_{\bm q}e^{i\bm q\cdot \bm r_{ij}}
\frac{1}{E-\xi_q^\Gamma\tau_z-\Delta\tau_x},
\end{equation}
with $\xi_q^\Gamma = \hbar^2q^2/2m_\Gamma-\mu_\Gamma$.

All Green's functions may be written in terms of integrals $I_0(E,\bm r_{ij})$ and $I_1(E,\bm r_{ij})$, which are analytically solvable for parabolic bands near the Fermi surface:
\begin{eqnarray}\label{greens}
G^K_{ij}(E) &=& 
\sum_{n\eta\lambda}\frac{e^{i\eta\bm K_n\cdot \bm r_{ij}}}{3}
[(E+\Delta\tau_x)I_0^{\eta\lambda}+\tau_zI_1^{\eta\lambda}
]
\frac{1+\lambda\sigma_z}{2},\notag\\
G^\Gamma_{ij}(E)&=&(E+\Delta\tau_x)I_0^\Gamma+\tau_z I_1^\Gamma.
\end{eqnarray}
The $I$ integrals are a sum of Bessel and Struve functions in agreement with other works on two-dimensional superconductors~\cite{Brydon2015, Heimes2015, Zhang2016} (see Appendix~\ref{app:int}).
Due to time-reversal symmetry of the pure superconductor, the integrals at $K$ valleys obey $I_{0,1}^{\eta\lambda} = I_{0,1}^{-\eta,-\lambda}$.
The equal-position total Green's function is trivial:
\begin{equation}\label{greenii}
G_{ii}(E)\simeq-\frac{\pi\rho(0)}{2}\frac{E+\Delta\tau_x}{\sqrt{\Delta^2-E^2}},
\end{equation}
with $\rho(0)$ the density of states at the Fermi surface for both spin projections.

\section{YSR states from a ferromagnetic chain of impurities}
\label{sec:ysr}
\subsection{General formulation}
In this section, we review the method to obtain the equation for the YSR bound states for a ferromagnetic chain of impurities.
Let us consider $N$ identical impurities coupled to the superconducting substrate and placed at equally distanced positions $\bm r_j$, with a spacing $d$.
By convention, the monolayer sits in $xy$ plane, and the direction of the linear chain is parametrized by the angle $\phi$ between the chain and $x$ axis.
Each impurity spin $\bm S$ is oriented along direction $\bm n$, specified by angles $\theta$ and $\varphi$, $\bm n=(\cos\varphi \sin\theta, \sin\varphi\sin\theta, \cos\theta)$.
We assume that in the following that the impurities are exchange-coupled identically to the substrate and the interaction is purely local in space.
Under these assumptions, the interaction Hamiltonian between the impurities and the substrate reads
\begin{equation}\label{h1}
H_1(\bm r)=-JS\sum_{j=1}^N \bm n\cdot\bm \sigma \delta(\bm r-\bm r_j),
\end{equation}
where $J>0$ is the exchange coupling constant between impurity and substrate, while $S$ is the impurity spin amplitude.
Within a classical approximation, the spins are treated as vectors, with amplitude $S \to \infty$, and small coupling constant, $J\to 0$, with magnetic exchange energy $JS$ finite.~\cite{Shiba1968}
Note that we neglect in the following the nonmagnetic interaction between impurities and substrate, since this potential scattering has typically a small influence on YSR state energy in $s$-wave superconductors.~\cite{Balatsky2006}

The preference for the ferromagnetic ordering in the dilute impurity limit has to be derived from a minimization of the indirect exchange Ruderman-Kittel-Kasuya-Yosida (RKKY) interaction energy between spins.~\cite{Ruderman1954, Kasuya1956, Yosida1957}
In Appendix~\ref{app:rkky}, we discuss the classical RKKY Hamiltonian showing that the Ising spin-orbit coupling induces, besides Heisenberg and Ising spin-spin interactions, also a Dzyaloshinskii-–Moriya interaction, which favor noncollinear ordering of impurity spins (similar to the case of \ce{MoS2}).~\cite{Parhizgar2013}
The results show that the RKKY interaction is direction dependent due to the $K$ valley electrons.
Therefore the ground state admits a variety of spin orders, depending on the distance between spins and chain orientation.
The analysis in Appendix~\ref{app:rkky} neglects additional crystal field effects which favors collinear order of spins~\cite{Heimes2014, Heimes2015} and could stabilize the spin order in Eq.~\eqref{h1}.
Therefore, at the moment, Eq.~\eqref{h1} remains an ansatz whose soundness requires additional numerical works or experimental investigations using spin-polarized STM.~\cite{Zhou2010, Khajetoorians2012, Nadj-Perge2014, Jeon2017, Jaeck2019}

The stationary Schr\"odinger equation reads:
\begin{equation}\label{schr}
[H_0(\bm r)+H_1(\bm r)]\psi_{\bm r}=E\psi_{\bm r}.
\end{equation}
Using the Fourier transform $\psi_{\bm r}=\frac{1}{\sqrt{\Omega}}\sum_{\bm k}\psi(\bm k)e^{i\bm k\cdot \bm r}$ and integrating over both sides in Eq.~\eqref{schr} with $\frac{1}{\sqrt{\Omega}}\int d\bm re^{-i\bm q\cdot \bm r}$ yields
\begin{equation}
H_0(\bm q)\psi(\bm q)-\frac{JS\bm n\cdot \bm \sigma}{\sqrt{\Omega}}\sum_j e^{-i\bm q\cdot\bm r_j}\psi_j=E\psi(\bm q).
\end{equation}
Using the free Green's function in momentum space $G(E,\bm q)^{-1}=E-H_0(\bm q)$, it follows the relation to determine $\psi(\bm q)$:
\begin{equation}
\psi(\bm q)=-\frac{JS}{\sqrt{\Omega}}\sum_jG(E,\bm q)e^{-i\bm q\cdot \bm r_j}\bm n\cdot\bm\sigma\psi_j.
\end{equation}
Integrating over momentum with $\frac{1}{\sqrt{\Omega}}\int d\bm q e^{i\bm q\cdot\bm r_i}$ and using the Green's function in real space, which for equidistant impurities depends only on the distance between impurities
located at $\bm r_i$ and $\bm r_j$,
\begin{equation}
G_{ij}(E)=\frac{1}{\Omega}\sum_{\bm q} G(E,\bm q)e^{i\bm q\cdot\bm r_{ij}},
\end{equation}
we obtain the equation for the YSR states:
\begin{equation}\label{eig}
\psi_i = -JS \sum_j G_{ij}(E)\bm n\cdot\bm \sigma\psi_j.
\end{equation}
The equation determines the wave function of YSR bound states localized at impurity positions, depending on the impurity spin orientations, the magnitude of the exchange coupling energy, and the substrate superconductor Green's function.
Due to the presence of the Green's function, it is however a nonlinear eigenvalue equation in the energy $E$.
To make analytical progress in determining the YSR wave functions and energies, it is necessary to perform additional approximations.

\subsection{Deep-impurity limit}
The YSR bound states may be obtained through a linearization of Eq.~\eqref{eig} in the so-called deep-impurity limit.
This approximation is correct when the YSR band is close to the Fermi level, when $|E|\ll |\Delta|$.
Such approximation is \textit{a priori} known to be physically applicable in the case of dilute impurities, where only weak interactions exist between neighboring impurity sites.
Under this condition, the picture from the single magnetic impurity physics is expected to hold, i.e.,~the YSR energy (and therefore the entire YSR band) may be brought at the Fermi level when free tuning the magnetic exchange energy $JS$ [see below Eq.~\eqref{classical}].
In the present section, we will therefore assume dilute weakly coupled impurities ($k_Fd\gg 1$) (for a dense-impurity approach, see Ref.~\onlinecite{Peng2015}) with energy $|E|\ll |\Delta|$.
Therefore one expands Eq.~\eqref{eig} both in the energetic distance of the bound states to the Fermi level and in a small coupling between impurity sites.
Keeping Eq.~\eqref{eig} at first order in the small parameters implies a linear dependence on energy in the same-position Green's function, while the Green's function controlling hopping is estimated at zero order in YSR energy:~\cite{Pientka2013}
\begin{equation}
G_{ij}(E) \simeq G_{ij}(0) + E\delta_{ij}\partial_E G_{ij}(0),
\end{equation}
where $\partial_E G_{ij}(0)$ denotes $\partial_E G_{ij}(E)|_{E=0}$ and $\delta_{ij}$ is the Kronecker delta.
Therefore a generalized eigenvalue equation follows from Eq.~\eqref{eig}:
\begin{equation}\label{hlin}
\sum_{j=1}^N[\delta_{ij}+G_{ij}(0)
JS\bm n\cdot\bm \sigma]\psi_j
=-E\partial_E G_{ii}(0)
JS\bm n\cdot\bm \sigma\psi_i.
\end{equation}
This result is further simplified with the use of Eqs.~\eqref{decomp} and \eqref{greenii} to cast the problem to a simple tight-binding stationary Schr\"odinger equation:
\begin{equation}
\sum_{j=1}^N H_{ij}\psi_j = E\psi_i,
\end{equation}
with the Hamiltonian 
\begin{eqnarray}
H_{ij}&=&\frac{\Delta}{\alpha}(\bm n\cdot\bm \sigma-\alpha\tau_x)\delta_{ij}\\
&&+\frac{JS\Delta}{\alpha}[G_{ij}^\Gamma(0)+(\bm n \cdot\bm\sigma)G_{ij}^K(0)(\bm n 
\cdot\bm \sigma)]
(1-\delta_{ij}),
\notag
\end{eqnarray}
with $\alpha = \pi \rho(0)JS/2$, and $\rho(0)$, the total density of states at the Fermi surface.
The on-site terms $(\delta_{ij})$ in the equation are explicitly separated from the hopping terms $(1-\delta_{ij})$, which depend on the Green's function for $\Gamma$ and $K$ valley electrons calculated before~\eqref{greens}.

If there is only one impurity added to the system, then the tight-binding equation contains only the on-site term:
\begin{equation}
E\psi = \frac{\Delta}{\alpha}(\bm n\cdot\bm \sigma - \alpha\tau_x)\psi.
\end{equation}
The YSR bound states follow readily, yielding the classical solutions~\cite{Balatsky2006}:
\begin{equation}\label{classical}
\frac{E}{\Delta} = \pm \frac{1-\alpha^2}{1+\alpha^2}.
\end{equation}
The two solutions cross each other at zero energy ($\alpha=1$) and a quantum phase transition takes place, resulting in a quasiparticle excited on the YSR state.
The two solutions are:
\begin{equation}
\psi_{+}=\frac12\pmat{1\\1} \otimes 
\pmat{\cos\frac{\theta}{2}\\ \sin\frac{\theta}{2}e^{i\varphi}}, 
\;
\psi_{-}=\frac12\pmat{1\\-1} \otimes 
\pmat{\sin\frac{\theta}{2}e^{-i\varphi}\\-\cos\frac{\theta}{2}}. 
\end{equation}

The tight-binding model simplifies when projecting it in the single impurity basis $\{\psi_{+},\psi_{-}\}$.
The resulting Hamiltonian has the same block structure as in Ref.~\onlinecite{Zhang2016}. 
Moreover, we recover their results if we conflate helicity and valley index and eliminate the contribution from the $\Gamma$ pocket (as it would be correct in gated \ce{MoS2} near the lowest point in the conduction band).
The effective Hamiltonian $\bar H$ reads:
\begin{equation}\label{tb_eff}
\bar H_{ij} = \pmat{
h_{ij}+b_{ij} & \bar\Delta_{ij} \\
-\bar\Delta_{ij}^* & -h_{ij}+b_{ij}
},
\end{equation}
with
\begin{eqnarray}
\frac{h_{ij}}{\Delta}&=&
\big(\frac{1}{\alpha}-1\big)\delta_{ij}+\frac{JS\Delta}{\alpha}\big[I_0^\Gamma(0,|\bm r_{ij}|)\notag\\
&&+\frac{1}{6}
\sum_{n\eta\lambda}e^{i\eta\bm K\cdot\bm r_{ij}} I_0^{\eta\lambda}(0,|\bm r_{ij}|)
\big](1-\delta_{ij}),\notag\\
\frac{b_{ij}}{\Delta}&=&\frac{JS\Delta}{6\alpha}\sum_{n\eta\lambda} \lambda
e^{i\eta\bm K\cdot\bm r_{ij}} I_0^{\eta\lambda}(0,|\bm r_{ij}|)\cos\theta,
\\
\frac{\bar\Delta_{ij}}{\Delta}
&=&-\frac{JS}{6\alpha}\sum_{n\eta\lambda}\lambda 
e^{i\eta\bm K_n \cdot\bm r_{ij}} I_1^{\eta\lambda}(0,|\bm r_{ij}|)
\sin\theta e^{-i\varphi}\notag.
\end{eqnarray}
The above Hamiltonian describes quasiparticle dynamics inside the YSR band, at subgap energies (note the scaling of all matrix elements with $\Delta$).
The terms proportional to $\lambda$ are due to spin-triplet correlations in the Green's function~\eqref{eq:gel}, and lead to antisymmetric ``effective magnetic field'' $b$ and effective superconducting pairing $\bar\Delta$, $b_{ij}=-b_{ji}$ and $\bar\Delta_{ij}=-\bar\Delta_{ji}$.
The antisymmetric pairing $\bar\Delta$ is reminiscent of the $p$-wave pairing in the Kitaev chain~\cite{Kitaev2001} and is essential to obtain topological superconductivity.
The integrals $I_{0,1}$ are expanded in the dilute limit $k_Fd\gg 1$, yielding the following asymptotic behavior as a function of distance between impurities (see Appendix~\ref{app:int}):
\begin{equation}
I_{0,1}(0, |\bm r_{ij}|)\sim \frac{1}{\sqrt{|\bm r_{ij}|}} e^{-|\bm r_{ij}|/\xi_0},
\end{equation}
with superconducting coherence length $\xi_0=\hbar v_F/\Delta$.
The slow power-law decay of $I_{0, 1}(0)$, up to distances on the order of $\xi_0\sim 700a$ in our system (see Appendix~\ref{app:FSparams}), results in a long-range hopping tight-binding model with direct hopping between sites at distance of $\mathcal O(\xi_0)$.

Two particular cases where there is no topological superconductivity may be directly inferred from the effective tight-binding Hamiltonian~\eqref{tb_eff}.
First, if the spins are entirely out-of-plane, then $\sin\theta=0$, and the effective gap $\bar\Delta$ closes in the model, making it impossible to have topological phases.
Second, the spin-orbit coupling which provides spin-triplet pairing is odd in momentum normal to the broken reflection symmetry axes ($x$ axis and its threefold rotations).
Therefore it must vanish along $y$ (and its threefold rotations), which leads to vanishing spin-orbit splitting along $\Gamma-M$ lines in the Brillouin zone [see Fig.~\ref{fig:fig1}(b)].
When the chain with direction given by $\bm r_{ij}$ sits along the $y$ axis or its threefold rotations, there is no contribution from triplet pairing and both $b$ and the effective gap $\bar\Delta$ vanish.
This leads again to a gapless trivial phase.
The above argument may be checked by writing $\bm K_n\cdot \bm r_{ij}=K|\bm r_{ij}|\cos(\frac{2\pi n}{3}-\phi)$, with the $K=4\pi/3a$, the distance from $\Gamma$ to $\pm\bm K_n$.
Hence, using the time-reversal symmetry property of $I$ integrals, $I^{\eta\lambda}_{0,1}=I^{-\eta,-\lambda}_{0,1}$, it follows that the effective magnetic field and superconducting gap are proportional to $\sum_{n=0}^2 \sin(K|\bm r_{ij}|\cos(\frac{2\pi n}{3}-\phi))$. 
Therefore they indeed vanish on the unbroken symmetry axes $\phi=(2m+1)\pi/6$, with $m$ an arbitrary integer.

The topological phases are determined by studying the Bloch Hamiltonian, obtained after a Fourier transform for momenta $k$ along the chain:
\begin{equation}
\bar H(k) = \sum_{j} e^{-ikjd} \bar H_j,
\end{equation}
with $jd$ the distance between impurities, with spacing $d$.
The effective Bloch Hamiltonian for the chain becomes:
\begin{equation}\label{bloch_eff}
\bar H(k) = \pmat{h(k) + b(k) & \bar\Delta(k) \\
\bar\Delta^*(k)	& -h(k) + b(k)
}
\end{equation}
with
\begin{eqnarray}
h(k) &=& \frac{1}{\alpha} -1 + \text{Re}[f_\Gamma(k)] +\frac{1}{3}\sum_{n\eta\lambda}
\textrm{Re}[f_{n\eta\lambda}(k)],\notag\\
b(k) &=& \frac{1}{3}\sum_{n\eta\lambda}\lambda\textrm{Re}[f_{n\eta\lambda}(k)]\cos\theta,\\
\bar\Delta(k) &=& \frac{1}{3}\sum_{n\eta\lambda}\lambda\textrm{Im}[f_{n\eta\lambda}(k)]\sin\theta e^{-i\varphi}
,\notag
\end{eqnarray}
and
\begin{widetext}
\begin{eqnarray}
f_{n\eta\lambda}(k) &=& -\frac{\rho_{\eta\lambda}(0)}{\rho(0)}
\sqrt{\frac{2}{\pi k_F^K d}}\left[
\textrm{Li}_{\frac12}\left(
e^{i(k-\eta K\cos(\frac{2\pi n}{3}-\phi)+k_{\eta\lambda}+i/\xi_0^K)d}
\right)
+
\textrm{Li}_{\frac12}\left(
e^{-i(k-\eta K\cos(\frac{2\pi n}{3}-\phi)-k_{\eta\lambda}-i/\xi_0^K)d}
\right)
\right]e^{-i\frac{\pi}{4}},\notag\\
f_{\Gamma}(k) &=& 
-\frac{\rho_{\Gamma}(0)}{\rho(0)}
\sqrt{\frac{2}{\pi k_F^\Gamma d}}\left[
\textrm{Li}_{\frac12}\left(
e^{i(k+k_F^\Gamma+i/\xi_0^\Gamma)d}
\right)
+
\textrm{Li}_{\frac12}\left(
e^{-i(k-k_F^\Gamma-i/\xi_0^\Gamma)d}
\right)
\right]e^{-i\frac{\pi}{4}}.
\end{eqnarray}
\end{widetext}
The expressions make use of the polylogarithm function 
$\textrm{Li}_\nu(z) = \sum_{j=1}^{\infty} z^j/j^\nu$.
The superconducting coherence length at pocket $x$ ($\Gamma$ or $K$) reads $\xi_0^x=\hbar v_F^x/\Delta$.
The values of Fermi momenta $k_F^x$, Fermi velocities $v_F^x$, and densities of states, at respective pockets, are explicit in Appendix~\ref{app:FSparams}.
Note that $\rho_\Gamma(0)$ is the total density of states at $\Gamma$, for both spin projections, while $\rho_{\eta\lambda}(0)$ is the density of states for valley $\eta$ and helicity $\lambda$.
Due to time-reversal invariance in pure \ce{NbSe2}, the following relation holds: $\rho_{\eta\lambda}(0)=\rho_{-\eta,-\lambda}(0)$.

\subsection{Topological phases}
Due to $C_3$ symmetry, all information about the Hamiltonian is contained for chain orientation angle $\phi$ spanning the interval $[0, \pi/3]$.
Moreover, due to unbroken in-plane reflection symmetry about $\pi/6$, there is the same physics at $\pi/6+\phi'$ and $\pi/6-\phi'$.
Therefore it is sufficient to study the effective Hamiltonian in the restricted window $[0, \pi/6]$, where we compute in the following the topological phase diagrams.

The effective Hamiltonian $\bar H$ has particle-hole symmetry represented by the operator $\tau_x\mathcal K$.
The presence of the magnetic field term $b(k)$ breaks time-reversal symmetry, and therefore the one-dimensional Hamiltonian is generally in D class,~\cite{Schnyder2008, Ryu2010} described by the topological number:
\begin{equation}\label{majnum}
\mathcal M = \sgn\{\textrm{Pf}[\tau_x \bar H(0)]\textrm{Pf}[\tau_x\bar H(\pi/d)]\} = \sgn[h(0)h(\pi/d)],
\end{equation}
where $\textrm{Pf}(A)$ is the Pfaffian of matrix $A$.
In numerics, we usually choose an in-plane ordering of impurity spins ($\theta=\pi/2$), since $b(k)$ vanishes in this case, ensuring that the YSR band is gapped.
Closing of the gap occurs at possible topological phase transitions or when the impurity chain sits along the in-plane unbroken symmetry lines ($y$ axis and its threefold rotations).
Note however that the case of in-plane spins is rather particular~\cite{Pientka2013, Pientka2014} because an effective time-reversal symmetry gets restored.
The Hamiltonian becomes real under a unitary transformation $U$ which gauges out the superconducting phase $e^{i\varphi}$:
\begin{equation}
U\bar H(\bm k) U^\dag = \tau_3 h(k)+\tau_1\bar\Delta(k)|_{\varphi=0},\; 
U = \pmat{e^{i\frac{\varphi}{2}} & 0\\ 0 & e^{-i\frac{\varphi}{2}}}.
\end{equation}
Consequently, the Hamiltonian is time-reversal symmetric under the action of $\tau_z\mathcal K$ and is classified in the one-dimensional BDI class.
In this case, the topological invariant is $\mathbb Z$,~\cite{Schnyder2008, Ryu2010} and multiple MBS are allowed in the model.
The presence of BDI phases in two-dimensional superconductors with in-gap YSR states have been studied in several other works where the substrate has Rashba spin-orbit coupling.~\cite{Pientka2013, Pientka2014, Heimes2014, Poeyhoenen2014}

\begin{figure}[t]
\includegraphics[width=\columnwidth]{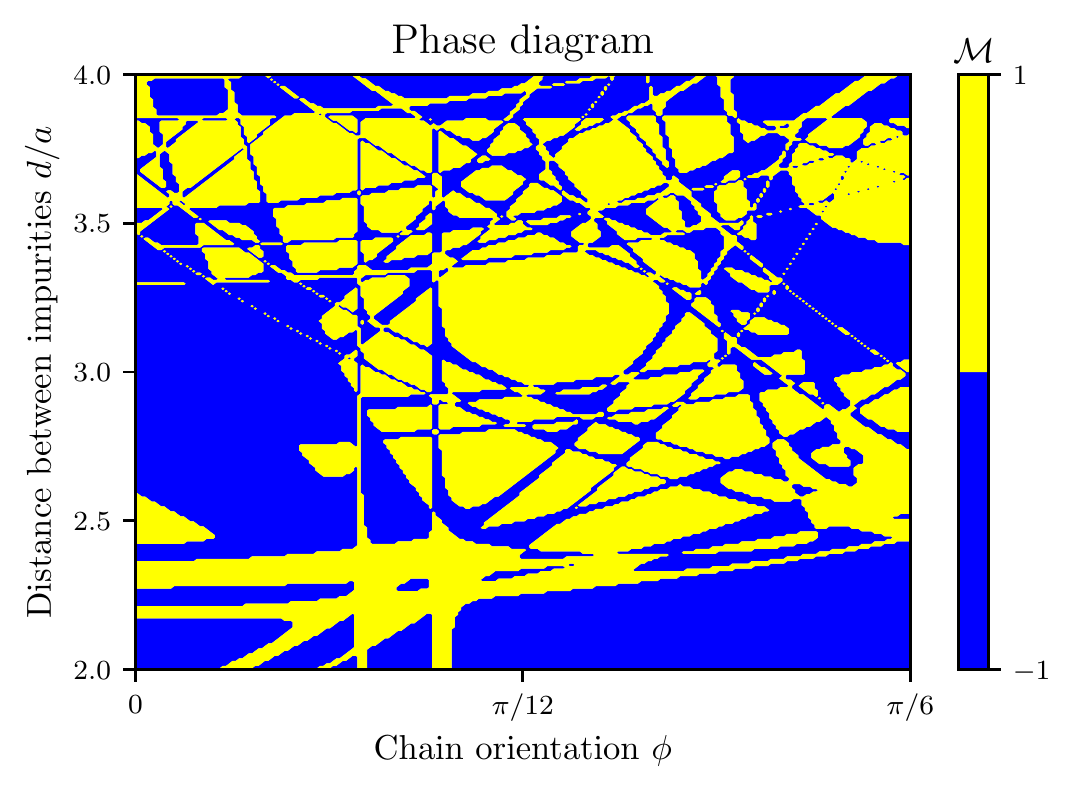}
\caption{Topological phase diagram as a function of the angle $\phi$ between the impurity chain and the $x$ axis and the distance between adjacent impurities $d$ in units of \ce{Nb} lattice constant $a$.
The magnetic exchange coupling is fixed at $JS=2/\pi\rho(0)$.
Blue regions denote the nontrivial topological phases with $\mathcal M=-1$.}
\label{fig:topdiag}
\end{figure}

Such behavior is not surprising and is apparent from symmetry considerations of the original Hamiltonian~\eqref{schr}.
In that case, the pure superconductor Hamiltonian $H_0$ is time-reversal invariant under the action of $T=i\sigma_y\mathcal K$.
The impurity spins contributing to $H_1$ act as a magnetic field breaking TRS.
If the impurity spins are constrained to point in plane, then they also break the out-of-plane mirror symmetry $M_z=-i\sigma_z$.
Nevertheless, the combination $M_zT=-i\sigma_x\mathcal K$ restores the correct transformation of spins under the time-reversal symmetry, placing the total Hamiltonian $H$ in the two-dimensional chiral-symmetric BDI class.
This conclusion holds more generally for a large class of TMDs including gated \ce{MoS2, MoSe2,} etc.~\cite{Wang2018}
Since the above argument is valid for any completely in-plane orientation of spins, all effective one-dimensional Hamiltonians derived from above and living in the plane, remain in BDI class.
Therefore, indeed, under this constraint on spins, the effective impurity chain Hamiltonian is in the one-dimensional BDI class.
 
In the present paper, we have limited ourselves to cases where only two MBS are present in the system, such that the Majorana number~\eqref{majnum} remains useful in characterizing the topological phases also for the chiral-symmetric BDI system.
For more generic cases of purely in-plane spins, the topological characterization requires the measurement of a winding number.~\cite{Tewari2012}
It is beyond the scope of the present study to inventory possible high-winding number phases of the purely planar orientation of spins. 
It is expected that generally there would be a nonvanishing out-of-plane component to spin projections, which renders high-winding number phases unlikely to be observed in experiments.
 
Figure~\ref{fig:topdiag} shows the resulting topological phase diagram as a function of distance between impurities and chain orientation $\varphi$. The magnetic exchange energy $JS$ is fixed at $2/\pi\rho(0)$, the point where the single-impurity YSR crosses zero, where we expect that the deep-impurity limit holds.
Despite the complicated structure of the phase diagram, several extended connected regions of nontrivial topological phases are present.
In particular, we will focus in the following numerical investigations on the region centered at $\phi=0$ and $d=3a$, and show that the topological phases are also robust when varying the magnetic exchange energy. 
Some additional results are contained in Appendix~\ref{app:topo}, when the distance between impurities is varied in a range relevant for the case of magnetic molecular chains with, e.g.,~phthalocyanine.

\section{Numerical results}
\label{sec:num_res}
In the following, we study the impurity problem using numerical methods, as a direct proof of MBS presence.
Section~\ref{sec:numdeep} investigates the deep-impurity limit in order to obtain the energy and wave functions of YSR states without resorting to the parabolic approximation of energy dispersion at the \ce{NbSe2} valleys.
Section~\ref{sec:exactdiag} dispenses with both the parabolic approximation and the deep-impurity limit by using exact diagonalization of a finite-size two-dimensional \ce{NbSe2} superconductor with impurities deposited on system sites.
All simulations manifest Majorana states at values predicted by the analytical study.

The numerical simulations come with one technical inconvenience. Due to the small superconducting gap $\Delta=\SI{1}{meV}$ in \ce{NbSe2}, the coherence lengths are on the order of $\xi_0\sim 700a$ (see Appendix~\ref{app:FSparams}).
To avoid unphysical finite size effects due to level spacing being larger that the superconducting gap, it is necessary for the tight-binding lattices to have linear lengths exceeding $\xi_0$.
This imposes a strain of numerical simulations, which is usually remedied by artificially increasing the superconducting pairing amplitude.
This increase does no modify qualitatively single-impurity physics since YSR energies are scaled by $\Delta$.
However, in the case of a chain of impurities, this has the effect to decrease the number of hopping terms involved in the effective Hamiltonian, and tends to favor the growth of topological phases size in parameter space (see Appendix~\ref{app:topo}).
Here we use in numerics $\Delta=\SI{10}{meV}$, i.e.,~a tenfold decrease of the coherence length.

\begin{figure}[ht]
\includegraphics[width=\columnwidth]{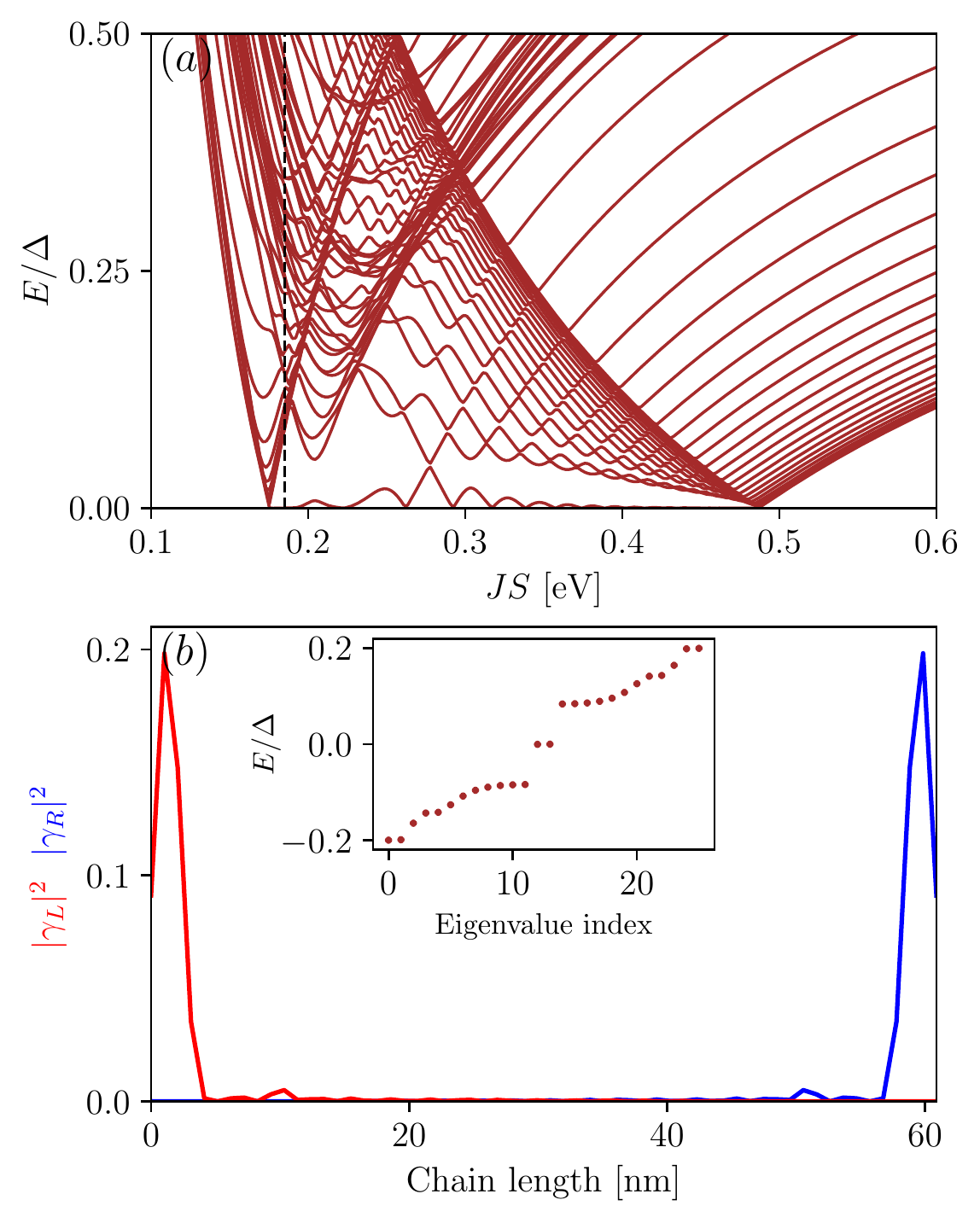}
\caption{Numerical results for a chain of 60 impurities on \ce{NbSe2} with distance between neighbor impurities $d=3a$, $a=\SI{0.344}{nm}$. (a) The deep-impurity limit YSR energies when varying the magnetic exchange energy $JS$. 
(b) For a fixed $JS=\SI{0.185}{eV}$, zero-energy Majorana modes appear in the middle of the gapped YSR band (inset) with wave function $\gamma_{L/R}$ strongly localized at the edges of the impurity chain (b).}
\label{fig:ej_lin}
\end{figure}

\subsection{Effective 1D models in the deep-impurity limit}
\label{sec:numdeep}
Here we obtain the YSR bands and MBS in the deep-impurity limit, without resorting to the parabolic approximation to the energy dispersion near the valleys.

The equal-position Green's function~\eqref{greenii} is trivial and its energy derivative follows analytically.
Therefore only left-hand side in Eq.~\eqref{hlin} needs to be computed numerically.
After denoting $\Xi_j=\bm\sigma\cdot\bm n\psi_j$, and rearranging the terms, the equation to determine YSR states becomes:
\begin{equation}
\frac{\Delta}{\alpha}\sum_{j=1}^N\left[\bm\sigma\cdot\bm n \delta_{ij}+JS G_{ij}(0)
\right]\Xi_j=E\Xi_i.
\end{equation}
The pure superconductor Green's function $G_{ij}(0)$ from Eq.~\eqref{tot_green} is calculated by numerical integration over the two-dimensional Brillouin zone.
We consider a periodic patch of sites defined by vectors $m\bm a_1$ and $n\bm a_2$, with $\bm a_1$ and $\bm a_2$ primitive vectors of the lattice, and $m, n$ integers.
This induces a discretization of the Brillouin zone, which contains $m\times n$ lattice momenta.
In numerical experiments, we usually choose $m=n=600$.

It was sufficient to consider small chains of 60 impurities, in order to have well-localized Majorana states.
The impurities sit at lattice positions and form a chain in the middle of the periodic patch.
For definiteness, we have considered a chain of impurities along $x$ axis ($\phi=0$) with spins pointing in the positive $x$ direction, spaced by $d=3a$. 
The results are shown in Fig.~\ref{fig:ej_lin} (upper panel), which explores the behavior of the YSR states and MBS in a range of magnetic exchange energies. 
There are two transition points where the YSR band closes, with an extended nontrivial topological phase in between, starting around $JS\sim \SI{0.2}{eV}$.
The topological phases are marked by the presence of oscillating MBS, separated by a gap from the rest of the YSR band.
The oscillating behavior of MBS is due to finite size effects, since oscillations decrease with the length of impurity chain and the size of the periodic substrate.
For a fixed value of $JS$, we present in Fig.~\ref{fig:ej_lin} (lower panel) the two MBS eigenvalues, and the amplitude of their functions, which is localized at the edges of the impurity chain.

\begin{figure}[t]
\includegraphics[width=\columnwidth]{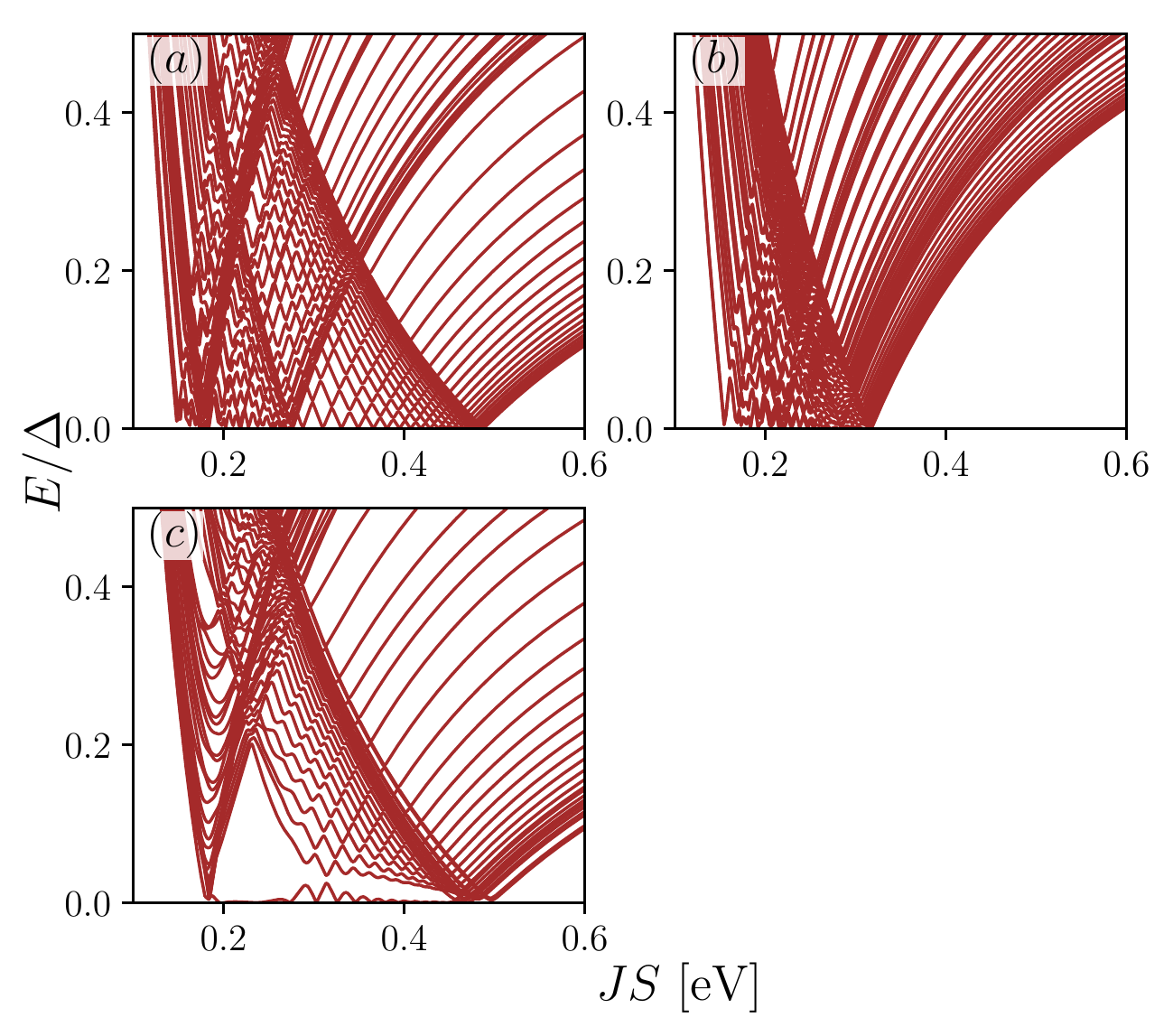}
\caption{Subgap YSR dispersion as a function of magnetic exchange energy $JS$. (a) Closing of the YSR band for purely out-of-plane projection of spins, with the chain oriented at $\phi=0$.
(b) Closing of the YSR band in the case of a chain of impurities placed along the conserved reflection symmetry axis at $\phi=\pi/6$, with impurities spaced by $d=2\sqrt 3 a$ and spins polarized in $x$ direction.
(c) MBS for some arbitrary orientation of the chain $(\phi\simeq 0.03)$ with distance between impurities $d\simeq 3.06a$ (the spins are polarized in $yz$ plane, with a small out-of-plane component fixed by an angle of $0.1$).
}
\label{fig:JS_add}
\end{figure}

The numerical simulations confirm the analytical prediction that the YSR gap closes for completely out-of-plane orientation of impurity spins Fig.~\ref{fig:JS_add}(a) and for impurity chains aligned along the conserved in-plane reflection symmetry axes Fig.~\ref{fig:JS_add}(b).
In both cases, no MBS develop in the system.
In contrast, Fig.~\ref{fig:JS_add}(c) shows a generic case with MBS for an arbitrary chain orientation, away from any symmetry axis of the triangular lattice, with an arbitrary orientation of impurity spins (with some nonzero out-of-plane component), and impurity positions incommensurate with the underlying lattice.

\subsection{Exact diagonalization of 2D tight-binding models}
\label{sec:exactdiag}

\begin{figure}[ht]
	\includegraphics[width=\columnwidth]{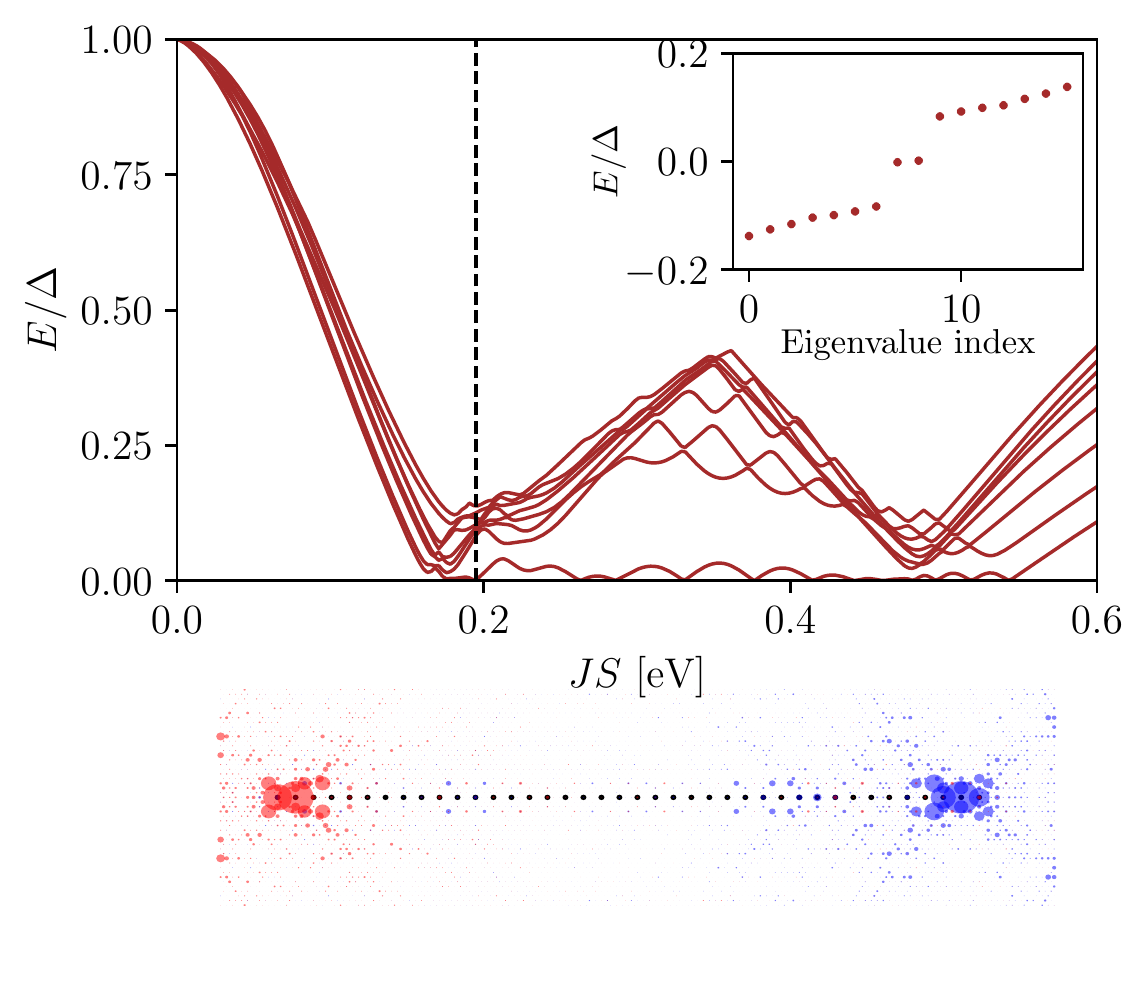}
	\caption{(Top) Lowest YSR energies from exact diagonalization of 2D lattices as a function of the magnetic exchange energy $JS$. For a fixed $JS=\SI{0.195}{eV}$ (dashed black line), a zoom shows in the inset the YSR gap with two zero-energy MBS.
	(Bottom) At the fixed $JS$, a cutout from the lattice shows the wave function localization of right (blue) and left (red) Majorana modes.
		The size of the blue and red dots is proportional to the absolute value squared of Majorana wave function, and therefore it presents a local density of states picture of the modes.
		The black dots mark the impurity positions in the simulation.}
	\label{fig:ed}
\end{figure}

In this section, MBS presence is proved through exact diagonalization of a finite triangular \ce{NbSe2} lattice.
To render simulations comparable to those of the previous section, we define impurities along $x$ axis, with spins polarized in the $x$ direction, and with distance between impurities $d = 3a$.
The lattice is a rectangle of size approximately $140a\times 70a$, such that it is twice longer than the estimated superconducting coherence length.
On the lattice, we fit a chain of 40 impurities arranged as in the lower panel of Fig.~\ref{fig:ed}, which shows a cutout (limited in $y$ direction) of the system.
A scan over different magnetic exchange energies shows MBS over a region comparable to that in the previous section.
Nevertheless, the finite size effects are more noticeable, compared to those in the larger lattices previously explored.
This leads to larger oscillations of MBS at zero energy and to a higher $JS$ where transition between trivial and nontrivial phases takes place.
We have checked by scaling up from smaller lattices, that the topological phases in $JS$ decreases, presumably towards the results shown in Fig.~\ref{fig:ej_lin}.
In the latter case of the previous section, we could check that the positions of topological phase transition do not change for larger lattices, but MBS oscillations are, as expected, reduced.

The inset of Fig.~\ref{fig:ed} shows the MBS eigenvalues, while the lower panel, the corresponding amplitudes of the wave functions for a fixed value of $JS$ in the nontrivial region.
The Majorana wave functions show strong localization at the edges of the impurity chains (lower panel in Fig.~\ref{fig:ed}).
Finite size effects are present in the slight overlap of MBS, leading to a small energy split near zero energy, as well as in the reflection of the MBS wave function from the vertical edges of the lattice (Fig.~\ref{fig:ed}, lower panel).
Simulations on the finite tight-binding systems have been performed using the \textsc{Kwant} package~\cite{Groth2014}.

\section{Discussion}
\label{sec:conc}
In this paper, we have studied topological phases supporting Majorana bound states due to ferromagnetic chains of impurities on monolayer \ce{NbSe2} superconductors.
Our analysis is based on realistic tight-binding modeling for the conduction band of the superconducting substrate using parameters extracted from fits to \textit{ab initio} band structure calculations.
Our results indicate that the strength of Ising spin-orbit coupling in the \ce{NbSe2} superconductor is sufficient to stabilize topological phases supporting Majorana bound states localized at the end of the impurity chains.
Similar to previous studies in transition metal dichalcogenides, such as \ce{MoS2},~\cite{Zhang2016, Sharma2016} we also showed that Majorana bound states are expected for generic orientation of impurity chains on the substrate and impurity spin projections.
Nevertheless, the Yu-Shiba-Rusinov bands close when the chains are oriented on the lines of unbroken in-plane reflection symmetry, rendering Majorana phases impossible in such cases.
Moreover, it is necessary for the existence of topological phases that impurity spins have a finite in-plane component.
Since topological phases in \ce{NbSe2} exist for spin-split bands at the $K$ valleys, it is reasonable to expect that topological phases would develop also in \ce{MoS2}, \ce{MoSe2}, etc. for generic gating in the conduction band, when two Fermi surfaces develop at each valley.

Our findings are encouraging for the recent experimental forays in generating YSR states in \ce{NbSe2}~\cite{Menard2015, Kezilebieke2018, Kezilebieke2019}.
In particular, a valid experimental work program would be to investigate whether self-assembled chains of magnetic molecules offers a robust alternative to designing one-dimensional topological superconductivity.
Nevertheless, additional studies are needed to determine the magnitude of magnetic exchange energies between the superconducting substrate and, e.g.,~magnetic porphyrin or phtalocyanine molecules.
Our work has determined that a necessary condition to generate nontrivial topological phases consists in having magnetic exchange energies of at least $JS\sim \SI{0.2}{eV}$.

Finally, a remaining open problem is to evaluate the effect of charge density wave ordering on the formation of the Yu-Shiba-Rusinov band and on the topological phases in \ce{NbSe2}.
Recent experiments~\cite{Liebhaber2019} demonstrate strong dependence of an YSR state energy on modulations in the local environment of the impurity.
Therefore we expect that the YSR band will be modified, and it remains to quantify the impact on topological superconducting phases.
\bibliographystyle{apsrev4-1}


\begin{acknowledgments}
We thank C.P.~Moca, F.~von Oppen, and L.P.~Z\^arbo, for illuminating discussions, and to M.P.~Nowak for going over the final manuscript.
This research was supported by a grant of the Romanian National Authority for Scientific Research  and  Innovation, CNCS-UEFISCDI, with project No.~PN-III-P1-1.2-PCCDI-2017-0338. 
\end{acknowledgments}

\appendix
\onecolumngrid
\section{Ab initio calculation}
\label{app:abinit}

The \textit{ab initio} simulations were performed using the \textsc{Siesta} code~\cite{Ordejon1996, Soler2002} which uses norm-conserving pseudopotentials~\cite{Troullier1991} and expands the wave functions of valence electrons using linear combinations of atomic orbitals (LCAO).
We used GGA/PBE.~\cite{Perdew1996} as exchange-correlation function.
The LCAO basis set is double-zeta polarized with an energy shift of $\SI{100}{meV}$.
The lattice constant is $\SI{0.344}{nm}$ and the integration in the Brillouin zone uses a $40\times 40$ Monkhorst-Pack grid.

The resulting band structure along the high-symmetry lines is included in Fig.~\ref{fig:fig1}(b).
The density of states for the conduction band is due almost entirely to $4p$ \ce{Se} and $4d$ \ce{Nb} orbitals.
The orbital-resolved density of states for the conduction band is given in Fig.~\ref{fig:dos_soc}(a).
Since the dominant contribution is due to \ce{Nb} atoms, it is justified to devise an effective model where the effect of \ce{Se} atoms is included in a Hamiltonian~\eqref{bloch} for hopping only in the triangular lattice of \ce{Nb} atoms.

\section{Low-energy tight-binding model}
\label{app:tb}
The explicit form for the Bloch-BdG Hamiltonian modeling monolayer \ce{NbSe2} from Eq.~\eqref{bloch} reads
\begin{eqnarray}\label{bloch_ham}
H_0(\bm k) &=& \xi_k\tau_z+\Lambda_{\bm k}\tau_z\sigma_z+\Delta\tau_x\notag\\
\xi_k &=& \mu 
+2t_1[\cos(2\alpha)+2\cos(\alpha)\cos(\beta)]
+2t_2[\cos(2\beta)+2\cos(3\alpha)\cos(\beta)]
+2t_3[\cos(4\alpha)+2\cos(2\alpha)\cos(2\beta)]
\notag\\ 
&&+4t_4[\cos(\alpha)\cos(3\beta)+\cos(4\alpha)\cos(2\beta)
+\cos(5\alpha)\cos(\beta)]
+2t_5[\cos(6\alpha)+2\cos(3\alpha)\cos(3\beta)],\\
\Lambda_{\bm k} &=&2\lambda_1[\sin(2\alpha)-2\sin(\alpha)\cos(\beta)]
+2\lambda_2[\sin(4\alpha)-2\sin(2\alpha)\cos(2\beta)])\notag,
\end{eqnarray}
where $\alpha = k_x a/2$ and $\beta = \sqrt{3}k_ya/2$.

The Hamiltonian parameters are obtained through a fit to the \textit{ab initio} data along the high-symmetry lines in BZ. The resulting parameters are presented in Table~\ref{tab:params}.
\begin{table}
\caption{Parameters (in \si{eV}) of the tight-binding Hamiltonian~\eqref{bloch} from a fitting of \textit{ab initio} data.}
\begin{tabular}{c c c c c c c c}
\hline\hline 
$\mu$ & $t_1$ & $t_2$ & $t_3$ & $t_4$ & $t_5$ & $\lambda_1$ & $\lambda_2$ \\
\hline
0.023 & 0.0134 & 0.097 & 0.0066 & -0.0102 & -0.0144 & 0.0163 & 0.0013 \\
\hline \hline
\end{tabular}
\label{tab:params}
\end{table}

\section{Derivation of the spin-orbit coupling term}
\label{app:soc}

\begin{figure}
\includegraphics[width=0.8\columnwidth]{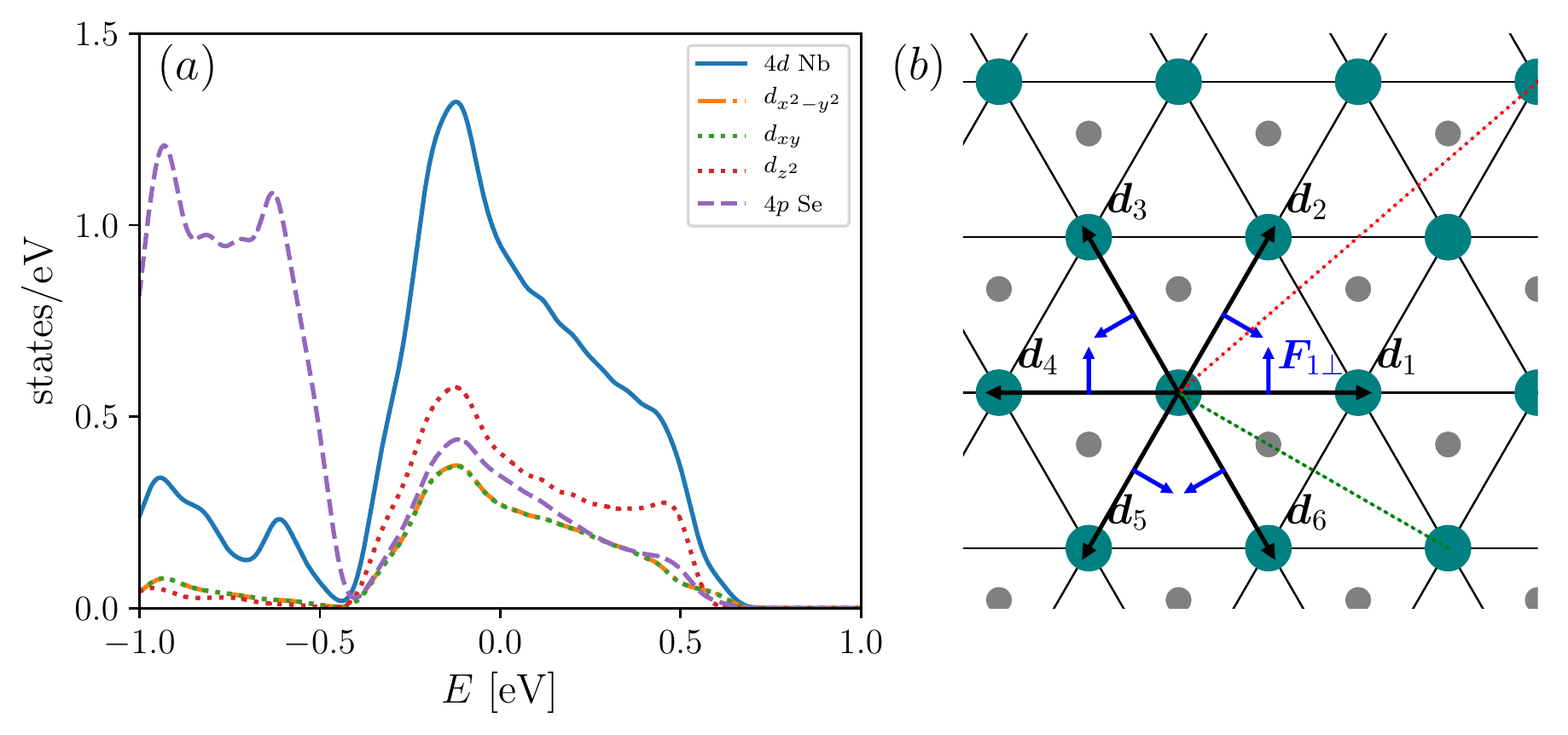}
\caption{
(a) Orbital-resolved density of states (per spin) in \ce{NbSe2}.
Only the orbital contribution relevant for the conduction band ($E_F=\SI{0}{eV}$) is shown, i.e.~the total contribution from $4d$ \ce{Nb} orbitals (continuous blue line) (due only to $d_{xy}$, $d_{x^2-y^2}, d_{z^2}$), and the $4p$ \ce{Se} orbitals (dashed purple line).
(b) The \ce{NbSe2} lattice viewed from the top.
The effective Hamiltonian~\eqref{bloch} models hopping between the \ce{Nb} atoms (teal). 
Nearest-neighbor hopping is along the $\bm d_j$ vectors.
The out-of-plane \ce{Se} (gray) atoms exert a force on electrons hopping along $\bm d_j$ (blue arrows, with only the force normal to $\bm d_1$, i.e.~$\bm F_{1\perp}$, being labeled).
The dashed green line corresponds to a second-nearest neighbor hopping, with no in-plane force exerted by \ce{Nb} or \ce{Se} atoms.
The dashed red line corresponds to fourth-nearest neighbor hopping, for which there exists only a small normal in-plane force (neglected by us in the modeling of tight-binding spin-orbit coupling) due to Coulomb repulsion  compensation from both sides of the hopping path.
}
\label{fig:dos_soc}
\end{figure}

An out-of-plane spin-orbit term is allowed due to breaking of the in-plane reflection symmetry $M_x=-i\sigma_x$. The spin-orbit coupling term $\Lambda_{\bm k}$ derives from the spin-orbit Hamiltonian for electrons on a lattice:
\begin{equation}
H_{\rm so} = \frac{\hbar}{4m^2c^2}(\bm\nabla V \times \bm p)\cdot\bm \sigma
= -\frac{\hbar}{4m^2c^2}(\bm F \times \bm p)\cdot\bm \sigma,
\end{equation}
where $m$ is the effective electron mass, $c$ the speed of light and $V$, the lattice potential.
The electronic cloud centered at $\ce{Se}$ atoms on top and at the bottom of the sheet of $\ce{Nb}$ atoms exert a Coulomb repulsion on electrons hopping in plane between the $\ce{Nb}$ atoms.
The effect is maximal for hopping along $\bm d_j$, when the force normal to the hopping is largest. 
For hopping along the other directions, the projection of $\ce{Se}$ atoms on the $\ce{Nb}$ plane passes close to the hopping path, such that there is compensation between the effect of the charges above and below the plane. 
For example, the compensation is perfect for next nearest-neighbor hopping, leading to a vanishing force [see Fig.~\ref{fig:dos_soc}(b)].
Under these condition, we investigate first the effect due to nearest-neighbor hopping.

Since the out-of-plane reflection symmetry $M_z = -i\sigma_z$ is preserved, the spin-orbit term may have only a component along $\sigma_z$.
Therefore the effect of spin-orbit term to nearest-neighbor hopping becomes in the second quantization:
\begin{equation}
H_{\rm so}^{(1)} = \frac{i}{2a}\frac{\hbar |\bm F_{\perp}|}{4 m^2c^2}
\sum_i\sum_{j=1}^6 \textrm{sgn}(\bm F_{j\perp}\times \bm d_j)c^\dag_{\bm r_i+\bm d_j}\sigma_z c_{\bm r_i}+\textrm{h.c.},
\end{equation}
where the sum over $i$ runs over all lattice sites and sum over spin indices is implied.
We used the symmetry to factor out the equal magnitude of forces $|\bm F_{\perp}|=|\bm F_{j\perp}|$ for any $j$.
Denoting the constant term $\lambda_1=\hbar |\bm F_{\perp}|/4m^2c^2 a$ and Fourier transforming into momentum space, we find a contribution from spin-orbit coupling:
\begin{equation}
H_{\rm so}^{(1)} =2\lambda_1\sum_{\bm k}
\left[
\sin(k_xa) - 
2\sin\left(\frac{k_xa}{2}\right)
\cos\left(\frac{\sqrt 3 k_ya}{2}\right)
\right]c_{\bm k}^\dag \sigma_z c_{\bm k}.
\end{equation}
Such spin-orbit coupling was used before~\cite{Xi2016, Sharma2016} to describe \ce{MoS2} type compounds modeled on a triangular lattice and may be derived generically for noncentrosymmetric superconductors~\cite{Samokhin2007}.
As noted, the contribution to next nearest-neighbor hopping vanishes. 
For fitting purposes, we also add the contribution of spin-orbit coupling to third nearest-neighbor hoppings, leading to the form that we used in our tight-binding Hamiltonian:
\begin{equation}
\Lambda_{\bm k} = 2\lambda_1[\sin(2\alpha)-2\sin(\alpha)\cos(\beta)]
+2\lambda_2[\sin(4\alpha)-2\sin(2\alpha)\cos(2\beta)],
\end{equation}
with $\alpha = k_x a/2$ and $\beta = \sqrt{3}k_ya/2$.
As expected, time-reversal symmetry of pure \ce{NbSe2}, together with the conserved reflection symmetry $M_y$, leads to vanishing spin-orbit coupling along $y$-axis (and its in-plane rotations under the threefold symmetry).

\section{Comparison with the three-band model}
\label{app:3bnd}
More elaborate models have been devised in order to reproduce the low-energy band structure of \ce{NbSe2} [as well as for other monolayer TMDs \ce{MX2}, with metal \ce{M} in \{\ce{Mo, W}\} and chalcogen \ce{X} in \{\ce{S, Se,Te}\}], the most used being a model for hopping only between three $d$ orbitals of \ce{Nb} in a triangular lattice.~\cite{Liu2013, He2018}
This model has more fitting parameters, allowing it to better reproduce the energy dispersion from the \textit{ab initio} data.
In addition, it also provides a fit for the first two bands above the Fermi level-crossing band.
In this appendix, we compare results between one-band \eqref{bloch} and three-band models in case of a single magnetic impurity in the system.

\begin{figure}
\includegraphics[width=0.8\columnwidth]{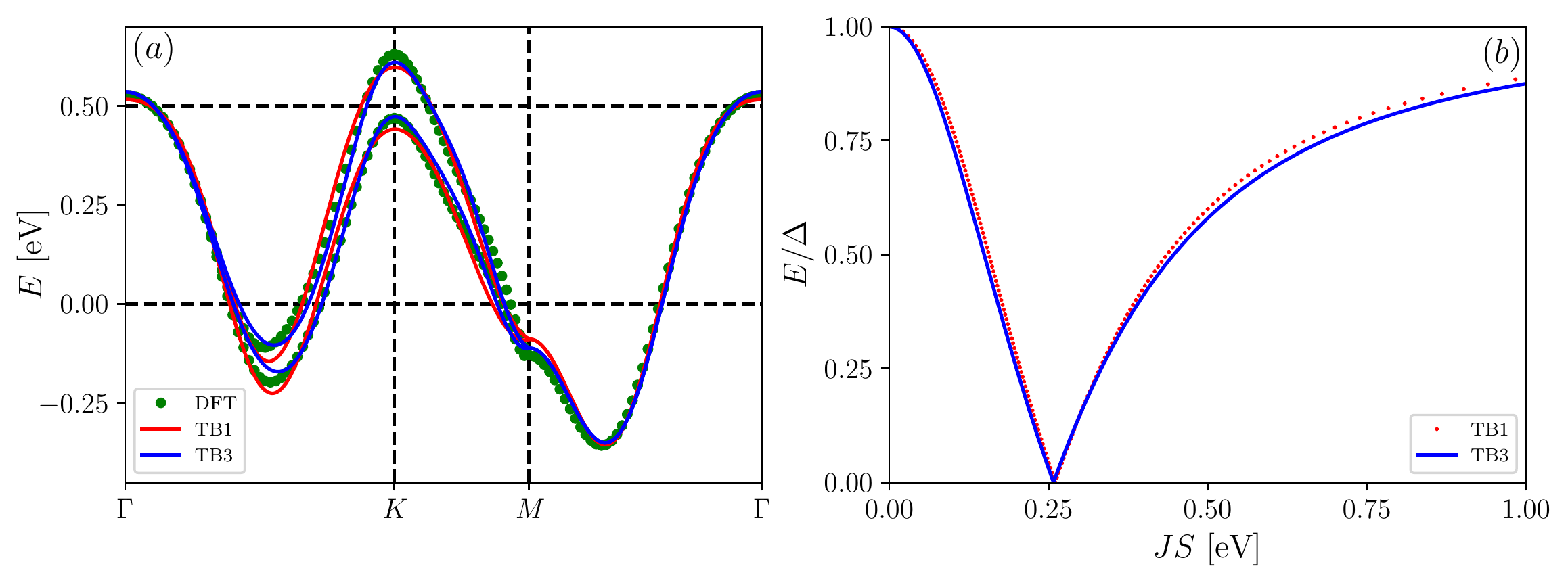}
\caption{(a) Energy dispersion for the single- (red, TB1, Eq.~\eqref{bloch_ham}) and three-band (blue, TB3, Eq.~\eqref{3bnd}) tight-binding models compared with the results from \textit{ab initio} simulations (green dots, DFT).
(b) Behavior of YSR energies for the case of a single impurity as the magnetic exchange energy varies in single- and three-band models.
}
\label{fig:3bnd_comp}
\end{figure}

The three-band Bloch Hamiltonian for superconducting \ce{NbSe2} reads
\begin{equation}\label{3bnd}
H_0(\bm k)=\tau_z\pmat{h_0 & h_1 & h_2\\
	h_1^* & h_{11} & h_{12} \\
	h_2^* & h_{12}^* & h_{22}
}
+\frac{\lambda}{2}\sigma_z
\pmat{0 & 0 & 0\\ 0 & 0 & -i\\ 0 & i & 0}
+\tau_y\sigma_y\Delta.
\end{equation}
The product of matrices is a Kronecker product, with Pauli matrices $\bm\tau$ in particle-hole space, and $\bm\sigma$ in spin space, with the 3-by-3 matrices in the orbital space. Absence of one matrix in a product is understood as presence of identity matrix in the respective space.
The matrix elements of the kinetic term are:
\begin{eqnarray}
h_0 &=& \varepsilon_1 + 2t_0 (2\cos \alpha\cos\beta+\cos2 \alpha)
+2r_0(2\cos 3\alpha\cos\beta+\cos2\beta
)
+2u_0(2\cos 2\alpha\cos 2\beta+\cos 4\alpha),
\notag\\
h_1 &=&2 i t_1 (\sin 2\alpha +\sin\alpha\cos\beta)
-2\sqrt{3}t_2\sin\alpha\sin\beta
+2(r_1+r_2)\sin 3\alpha\sin\beta
-2\sqrt{3}u_2\sin 2\alpha\sin 2\beta\notag\\
&&+2i(r_1-r_2)\sin 3\alpha\cos\beta
+2i u_1\sin 2\alpha(2\cos2\alpha+\cos2\beta),
\notag\\
h_2&=&2 i \sqrt3 t_1 \cos\alpha\sin\beta +
2t_2(\cos 2\alpha-\cos\alpha\cos\beta)
-\frac{2}{\sqrt3}(r_1+r_2)(\cos3\alpha\cos\beta-\cos 2\beta)
+2u_2(\cos4\alpha-\cos2\alpha\cos2\beta)
\notag\\
&&+\frac{2i}{\sqrt 3}(r_1-r_2)\sin\beta(\cos 3\alpha+2\cos\beta)
+2i\sqrt{3}u_1\cos 2\alpha\sin 2\beta,
\\
h_{11}&=&\varepsilon_2 + t_{11} (\cos\alpha\cos\beta+2 \cos2 \alpha)
+3 t_{22} \cos\alpha \cos\beta
+4r_{11}\cos 3\alpha\cos\beta
+2(r_{11}+\sqrt 3 r_{12})\cos 2\beta
\notag\\
&&+(u_{11}+3u_{22})\cos 2\alpha\cos 2\beta
+2u_{11}\cos 4\alpha,
\notag\\
h_{22}&=&\varepsilon_2 + 3t_{11} \cos\alpha\cos\beta
+t_{22}(\cos\alpha\cos\beta +2\cos2\alpha )
+2r_{11}(2\cos3\alpha\cos\beta+\cos2\beta)
+\frac{2}{\sqrt 3}r_{12}(4\cos3\alpha\cos\beta-\cos2\beta)\notag\\
&&+(3u_{11}+u_{22})\cos2\alpha\cos2\beta+2u_{22}\cos4\alpha,
\notag\\
h_{12}&=&\sqrt3 (t_{22}-t_{11})\sin\alpha\sin\beta
+4 i t_{12} \sin\alpha(\cos\alpha-\cos\beta)
+4r_{12}\sin3\alpha\sin\beta+\sqrt3(u_{22}-u_{11})\sin2\alpha\sin2\beta
\notag\\
&&+4iu_{12}\sin2\alpha(\cos2\alpha-\cos2\beta),
\notag
\end{eqnarray}
where $\alpha=k_xa/2$ and $\beta=\sqrt 3 k_ya/2$.
With the three-band model, the fit reproduces the conduction band together with the first two higher bands coming from the \textit{ab initio} calculations.
Note that the model may be used to fit perfectly only the conduction band, at the expense of a poor fit for the higher energy bands.
Instead, we fit all the three above mentioned bands, which leads to small deviations of the fitted model in the conduction band, mostly at the minimum between $\Gamma$ and $K$ valleys.
The resulting fitting parameters are shown in Table~\ref{tab:3bnd}, together with the energy dispersion along the high-symmetry lines, in Fig.~\ref{fig:3bnd_comp}(a).

To obtain YSR states for a single impurity, we first connect a site in the substrate to one impurity. 
The classical spin orientation is irrelevant to bound state energy in case of a single impurity and we fix it here normal to the monolayer. 
From the three-band model, we extract the density of states at the Fermi surface ($\rho(0)\simeq 2.468$ states$/\si{eV}$) and use Eq.~\eqref{classical} to determine the YSR energies as a function of $JS$ in Fig.~\ref{fig:3bnd_comp}(b).
In the one-band case we perform the calculation of YSR energies numerically, using the methods described in Sec.~\ref{sec:numdeep}.
The results are remarkably similar due to only a small variation in the Fermi surface density of states between the two models.
They both identify the crossing of YSR states at zero energy at $JS\approx \SI{0.25}{eV}$.
These results are encouraging for the focus in our paper on the simpler one-band model, which is amenable to analytical treatment in the parabolic approximation, and provides a less CPU time-consuming numerical implementation.

\begin{table}
\caption{Parameters (in \si{eV}) of the three-band tight-binding Hamiltonian~\eqref{3bnd} from a fitting of \textit{ab initio} data.}
\begin{tabular}{c c c c c c c c c c c}
\hline\hline $t_0$ & $t_1$ & $t_2$ & $t_{11}$ & $t_{12}$ & $t_{22}$ & $\varepsilon_1$ & $\varepsilon_2$ & $\lambda$ \\
$u_0$ & $u_1$ & $u_2$ & $u_{11}$ & $u_{12}$ & $u_{22}$ & $r_0$ & $r_1$ & $r_2$ & $r_{11}$ & $r_{12}$ \\
\hline
$-0.2399$ & 0.3245 & 0.3389 & 0.2608 & 0.2918 & $-0.0511$ & $1.5047$ & 1.9518 & 0.1372 \\
0.0686 & $-0.0421$ & 0.0105 & 0.0919 & $-0.0254$ & $-0.1127$ & 0.0098 
 & $-0.0727$ & 0.0101 & 0.0365 & 0.0871
\\
\hline \hline
\end{tabular}
\label{tab:3bnd}
\end{table}

\section{Parameters in the analytical calculation}
\label{app:FSparams}
The only free parameters in the analytical calculation are the chain orientation on top of the two-dimensional superconductor, given by the angle $\phi$, the orientation of the classical impurity spins, the distance $d$ between adjacent impurities, and the amplitude of magnetic exchange energy $JS$.
The other constants are fixed from the tight-binding model and is extracted in an investigation of the energy dispersion at the $\pm K$ and $\Gamma$ pockets.

Near the $\Gamma$ pocket, the Fermi surface is close to spherical. From a fit of the analytical model to the \textit{ab initio} data, we obtain:
\begin{equation}
k_F^\Gamma a \simeq 1.623, \quad
\hbar v_F^\Gamma /a \simeq \SI{0.64}{eV}.
\end{equation}

Near the $\pm K$ pockets, the surface is hexagonally warped even in the absence of spin-orbit coupling.
To obtain constants appropriate for a spherically symmetric effective Hamiltonian at the pockets, we perform the angular average of the Fermi surface $\bar x_F = \frac{1}{2\pi}\int d\chi x_F(\chi)$, with $x$ either Fermi velocity or momentum.
This yields effective Fermi momenta and velocities (omitting the bar):
\begin{equation}
k_F^K a\simeq 1.43,\quad \hbar v_F^K/a \simeq \SI{0.73}{eV}.
\end{equation}
From the split of the Fermi sheets at the $K$ point, we extract the spin-orbit momentum:
\begin{equation}
k_{so}a\simeq 0.1.
\end{equation}
The above results imply that the superconducting coherence length is on the order of $\xi_0\sim \hbar v_F/\Delta \sim 700 a$ with $a = \SI{0.344}{nm}$.

Using numerical integration at the Fermi surface with a broadening of the Green's function $\epsilon=10^{-3}$, we obtain the densities of states:
\begin{eqnarray}\label{rho_params}
\rho_\Gamma(0) &\simeq&  \SI{0.75}/ \si{eV}\Omega_0,\notag\\
\rho_{++}(0) &=&\rho_{--}(0)\simeq 0.4/\si{eV}\Omega_0,\notag\\
\rho_{+-}(0) &=&\rho_{-+}(0)\simeq 0.38/\si{eV}\Omega_0,\\
\rho(0) &\simeq& 2.32/\si{eV}\Omega_0,\notag
\end{eqnarray}
in states per electronvolt and primitive unit cell area $\Omega_0=\sqrt3 a^2/2$.
Note that $\rho_\Gamma(0)$ and $\rho(0)$ are total densities of states for both spin projections, and $\rho_{\eta\lambda}(0)$, is the density of states at $K$ valley $\eta=\pm$ for helicity $\lambda=\pm$.

\section{Integrals in the analytical model}
\label{app:int}
The integrals in Eq.~\eqref{greens} are
\begin{equation}
I_0^{\eta\lambda}(E, \bm r_{ij})=\frac{1}{\Omega}\sum_{\bm q}e^{i\bm q\cdot \bm r_{ij}}
\frac{1}{E^2-\xi_{q, \eta\lambda}^2-\Delta^2},
\end{equation}
\begin{equation}
I_1^{\eta\lambda}(E, \bm r_{ij})=\frac{1}{\Omega}\sum_{\bm q}e^{i\bm q\cdot\bm r_{ij}}
\frac{\xi_{q,\eta\lambda}}{E^2-\xi^2_{q, \eta\lambda}-\Delta^2},
\end{equation}
in the limit of many momenta $\bm q$ at the Fermi surface for the $K$ pockets. 
At energies $|E|<|\Delta|$ much smaller than the normal metal bandwidth, we assume that the density of states varies slowly, such that the following approximation holds $\frac{1}{\Omega}\sum_{\bm q} = \frac{\rho_{\eta\lambda}(0)}{2\pi}\int_0^{2\pi} d\vartheta\int d\xi$, with $\rho_{\eta\lambda}(0)$ the density of states at the Fermi level for the Fermi surface at valley $\eta$ and helicity $\lambda$.
In this approximation the momenta are linearized at the Fermi surface,
$\bm q\cdot \bm r_{ij} = q(\xi)|\bm r_{ij}|\cos\vartheta$, with 
\begin{equation}
q(\xi)\simeq k_{\eta\lambda}+\xi/\hbar v_F^K, \quad k_{\eta\lambda}=k_F^K - \eta\lambda k_{\rm so}.
\end{equation}
The last approximation assumes a parabolic band instead of the warped surface seen in experiments or \textit{ab initio} simulations.
In order to make our calculations realistic, we extract an effective (parabolic band) Fermi momentum $k_F^K$ and effective velocity $v_F^K$ from the tight-binding model by performing an angular average over the warped Fermi surface (see Appendix~\ref{app:FSparams}).

Under these approximations, the $I_0$ integral is well-behaved at the Fermi surface and readily yields the result:
\begin{equation}
I_0^{\eta\lambda}(E, |\bm r_{ij}|) = -\frac{\pi\rho_{\eta\lambda}(0)}{\sqrt{\Delta^2-E^2}}
\textrm{Re}[J_0(z_{\eta\lambda}^{ij})+iH_0(z_{\eta\lambda}^{ij})],
\end{equation}
with $J_0$, the zero order Bessel function and $H_0$, the zero order Struve function.
Their argument is
\begin{equation}
z_{\eta\lambda}^{ij}=(k_{\eta\lambda} +i/\xi_E^K)|\bm r_{ij}|,
\end{equation}
with superconducting coherence length $\xi_E^K=\hbar v_F^K/\sqrt{\Delta^2-E^2}$.

The $I_1$ integral has logarithmic divergences at high energies, which are resolved by introducing a regularization, which ensures convergence of the integral at energies higher than Debye energy $\omega_D$~\cite{Pientka2013}:
\begin{eqnarray}
I_1^{\eta\lambda}(E, |\bm r_{ij}|) & \simeq & \frac{\rho_{\eta\lambda}(0)}{2\pi}
\int_{-\infty}^\infty d\xi \frac{\omega_D^2}{\omega_D^2+\xi^2} \int_0^{2\pi} d\vartheta
e^{iq(\xi)|\bm r_{ij}|\cos\vartheta} \frac{\xi}{E^2-\xi^2-\Delta^2},
\\
&=& \pi\rho_{\eta\lambda}(0)
\textrm{Im}[J_0(z_{\eta\lambda}^{ij})+iH_0(z_{\eta\lambda}^{ij})].\notag
\end{eqnarray}

In the limit of dilute impurities $k_F d\gg 1$, the asymptotic expansion of Bessel and Struve functions in $I_{0,1}$ yields:
\begin{eqnarray}
I_0^{\eta\lambda}(E, |\bm r_{ij}|)&\simeq& - \frac{\pi\rho_{\eta\lambda}(0)}{\sqrt{\Delta^2-E^2}}
\sqrt{\frac{2}{\pi k_F^K |\bm r_{ij}|}}\cos(k_{\eta\lambda}|\bm r_{ij}|-\frac{\pi}{4})
e^{-|\bm r_{ij}|/\xi_E^K}\notag\\
I_{1}^{\eta\lambda}(E, |\bm r_{ij}|)&\simeq&
\pi\rho_{\eta\lambda}(0)
\sqrt{\frac{2}{\pi k_F^K |\bm r_{ij}|}}\sin(k_{\eta\lambda}|\bm r_{ij}|-\frac{\pi}{4})
e^{-|\bm r_{ij}|/\xi_E^K}.
\end{eqnarray}
The integrals at the $\Gamma$ point are solved identically, and the results follow from above through the replacements $k_{\eta\lambda}\to k_F^\Gamma$, $v_F^K\to v_F^\Gamma$, $\xi_E^K\to \xi_E^\Gamma$, and $\rho_{\eta\lambda}(0)\to\rho_\Gamma(0)/2$.

\section{RKKY Hamiltonian and competing spin-spin interactions}\label{app:rkky}
In this Appendix, we analyze in the parabolic approximation of Sec.~\ref{sec:model} the RKKY Hamiltonian which is responsible for the ordering of impurity spins on the metallic \ce{NbSe2}.
To obtain the Hamiltonian, we will work in the limit of dilute impurities $k_F r\gg 1$ and we assume that the ordering of impurity spins is set prior to \ce{NbSe2} becoming superconducting.
Our analysis follows similar works~\cite{Heimes2015} by using Matsubara Green's function formalism to single out the dominant contribution from conduction electrons near the Fermi surface.

Let us assume each impurity coupled to the surface through a Hamiltonian:
\begin{equation}
H_1(\bm r)=-J\sum_{j=1}^N \bm S_j\cdot\bm \sigma \delta(\bm r-\bm r_j),
\end{equation}
where in contrast to Eq.~\eqref{h1}, there is no preference for a particular ordering of spins $\bm S_j$.
Since $\bm \sigma$ are the Pauli spin matrices, and $\bm S_j$ are classical spins at position $\bm r_j$, the coupling constant $J$ has units (length)$^2/$(time).
The spins order through indirect interactions mediated by the conduction electrons, i.e.~the RKKY mechanism.

A second order perturbation theory yields the RKKY Hamiltonian,~\cite{Ruderman1954,Kasuya1956,Yosida1957} which in the zero temperature limit reads:
\begin{equation}\label{h_rkky}
H_{\rm RKKY} = \frac{J^2}{2}\sum_{jl}\int_{-\infty}^{\infty} \frac{d\omega}{2\pi}\textrm{Tr}[(\bm S_j\cdot \bm \sigma)g_{jl}(i\omega) 
(\bm S_l\cdot \bm \sigma)g_{lj}(i\omega)].
\end{equation}
The prefactor $1/2$ is added to avoid double counting $(jl)$ pairs.
The Matsubara Green's function  $g_{jl}(i\omega)$ for normal-state metallic \ce{NbSe2} is evaluated at the Fermi surface where it can be decomposed into separate contributions from $\Gamma$ and $K$ electrons.
In direct parallel to Eqs.~(\ref{tot_green}-\ref{GG}) in the main text, we find the Green's functions:
\begin{eqnarray}
g_{jl}^\Gamma(i\omega) &=& \frac{1}{\Omega}\sum_{\bm q}
e^{i\bm q\cdot\bm r_{jl}}\frac{1}{i\omega-\xi_q^\Gamma},\notag\\
g_{jl}^K(i\omega)&=& \frac{1}{3\Omega}
\sum_{n\eta\lambda,\bm q}
e^{i(\eta\bm K_n+\bm q)\cdot\bm r_{jl}} g^{\eta\lambda}(i\omega,\bm q),\\
g^{\eta\lambda}(i\omega,\bm q)&=&\frac{1}{i\omega-\xi_{q,\eta\lambda}}\frac{1+\lambda\sigma_z}{2}.
\notag
\end{eqnarray}
The integrals over momenta are performed formally following the same route as in the previous appendix.
The summation of momenta in the limit of many $\bm q$ becomes an integral.
Focusing on an small energy interval near the Fermi surface, the integrals over momenta may be approximated using the constant density of states at the Fermi surface, $\frac{1}{\Omega}\sum_{\bm q}\to \frac{\rho(0)}{2\pi}\int d\vartheta\int d\xi$.
To simplify the ensuing expressions, we assume only in this appendix that the densities of states for different helicities are identical $\rho_{\eta\lambda}(0)\simeq \rho_{\eta,-\lambda}(0)\simeq \rho_K(0)/2$, with $\rho_K(0)\simeq 0.78/\si{eV}\Omega_0$ the total density of states at a given valley. 
This approximation is reasonable since $\rho_{++}$ is only $5\%$ larger than $\rho_{+-}$ [see Eq.~\eqref{rho_params}].
The qualitative effects due to spin-orbit coupling enter through the oscillating exponential factors in the Green's functions.
Under these conditions, the angular integrals follow immediately, yielding a first kind zero order Bessel function $J_0(q(\xi)r)$.
In the limit of dilute impurities $k_Fr\gg 1$, we use the asymptotic expansion of the Bessel function for large argument:
\begin{equation}
J_0(z)\sim \sqrt{\frac{2}{\pi z}}\cos(z-\pi/4).
\end{equation}
The prefactor in the asymptotic expansion of $J_0$ is evaluated at Fermi surface, neglecting also a small spin-orbit coupling $k_F\gg k_{so}$. 
Then performing the last integral over $\xi$ yields the following Matsubara Green's functions at $K$ and $\Gamma$:
\begin{eqnarray}
g^\Gamma_{jl}(i\omega) &=&
-i\textrm{sgn}(\omega)\sqrt{\frac{\pi}{2k_F^\Gamma r_{ij}}}\rho_\Gamma(0)
e^{i\textrm{sgn}(\omega)(k_F^\Gamma-\pi/4)r_{jl}}e^{-\frac{|\omega|r_{jl}}{\hbar v^\Gamma_F}},
\notag\\
g^K_{jl}(i\omega) &=& \sum_{n\eta\lambda}e^{i\eta \bm K_n\cdot \bm r_{jl}}I_{jl}^{\eta\lambda}(i\omega)\frac{1+\lambda\sigma_z}{2},\\
I_{jl}^{\eta\lambda}(i\omega) &=& -i\textrm{sgn}(\omega)\sqrt{\frac{\pi}{2k_F^K r_{jl}}}\rho_K(0)
e^{i\textrm{sgn}(\omega)(k_{\eta\lambda}-\pi/4)r_{jl}}e^{-\frac{|\omega|r_{jl}}{\hbar v^K_F}}.\notag
\end{eqnarray}
We have used the same notations as before for Fermi velocities, Fermi momenta, densities of states, and $k_{\eta\lambda}=k_F^K-\eta\lambda k_{so}$.
Note that $g_{jl}^\Gamma(i\omega)$ and $I_{jl}^{\eta\lambda}(i\omega)$ depend only on the absolute distance $r_{jl}=|\bm r_{jl}|$, so they are even under permutation of real-space indices.

Let us return into the RKKY Hamiltonian~\eqref{h_rkky}, using the Green's function decomposition at the Fermi surface.
The ensuing Hamiltonian contains three types of contributions. 
The interaction between the spin at $j$ and $l$ is mediated entirely through conduction electrons at $\Gamma$ valley, or at $K$ valley, or through both $\Gamma$ and $K$ valleys. 
Therefore the Hamiltonian comprises three possible integrals over Green's functions:
\begin{eqnarray}\label{ints_mats}
\int\frac{d\omega}{2\pi}[g^\Gamma_{jl}(i\omega)]^2
&=&-\frac{\rho_{\Gamma}(0)}{4\pi r_{jl}^2}\sin(2 k_F^\Gamma r_{jl}),\notag\\
\int\frac{d\omega}{2\pi}I^{\eta\lambda}_{jl}(i\omega)I^{\eta'\lambda'}_{jl}(i\omega)
&=&-\frac{\rho_K(0)}{4\pi r_{jl}^2}\sin((k_{\eta\lambda}+k_{\eta'\lambda'})r_{jl}),\\
\int\frac{d\omega}{2\pi}g^\Gamma_{jl}(i\omega)I^{\eta\lambda}_{jl}(i\omega)
&=&-\frac{\rho_\Gamma(0)\rho_K(0)}{2\sqrt{k_F^K k_F^\Gamma}r_{jl}^2}
\frac{\hbar v_F^\Gamma v_F^K}{v_F^\Gamma+v_F^K}\sin((k_F^\Gamma+k_{\eta\lambda})r_{jl}).\notag
\end{eqnarray}
Subsequently, the trace over spins is performed using the identities:
\begin{eqnarray}\label{spin_traces}
\textrm{Tr}[(\bm\sigma\cdot \bm S_j)(\bm\sigma\cdot \bm S_l)] &=& 2\bm S_j\cdot\bm S_l,\notag\\
\textrm{Tr}[(\bm\sigma\cdot \bm S_j)(\bm\sigma\cdot \bm S_l)\sigma_z] 
&=& -\textrm{Tr}[(\bm\sigma\cdot \bm S_j)\sigma_z(\bm\sigma\cdot \bm S_l)] = 2i (\bm S_j\times\bm S_l)_z
\\
\textrm{Tr}[(\bm\sigma\cdot \bm S_j)\sigma_z(\bm\sigma\cdot \bm S_l)\sigma_z] 
&=& 4 S^z_jS^z_l-2\bm S_j\cdot\bm S_l.\notag
\end{eqnarray}
Combining Eqs.~\eqref{ints_mats} and~\eqref{spin_traces} in the RKKY Hamiltonian~\eqref{h_rkky} yields:
\begin{eqnarray}
H_{\rm RKKY} &=& -\frac{J^2}{2}\sum_{jl}\frac{1}{r_{jl}^2}\bigg\{
\frac{\rho_\Gamma(0)}{2\pi} \bm S_j\cdot\bm S_l\sin(2k_F^\Gamma r_{jl}) \\
&&+\frac{\rho_\Gamma(0)\rho_K(0)}{3\sqrt{k_F^K k_F^\Gamma}}
\frac{\hbar v_F^K v_F^\Gamma}{v_F^K+v_F^\Gamma}
\sum_{n\eta\lambda}
\bigg[\cos(\bm K_n\cdot\bm r_{jl})(\bm S_i\cdot\bm S_j)
+\lambda\eta\sin(\bm K_n\cdot\bm r_{jl})(\bm S_j\times \bm S_l)
\bigg]
\sin((k_{\eta\lambda}+k_F^\Gamma)r_{jl})\notag\\
&&+\frac{\rho_K(0)}{72\pi}\sum_{\substack{n\eta\lambda\\n'\eta'\lambda'}}
\bigg[\cos((\eta\bm K_n-\eta'\bm K_{n'})\cdot \bm r_{jl})
((1-\lambda\lambda')\bm S_j\cdot\bm S_l+2\lambda\lambda' S_j^zS_l^z)\notag\\
&&+(\lambda-\lambda')\sin((\eta\bm K_n-\eta'\bm K_{n'})\cdot \bm r_{jl})(\bm S_j\times\bm S_l)_z
\bigg]\sin((k_{\eta\lambda}+k_{\eta'\lambda'})r_{jl})
\bigg\}.
\end{eqnarray}
The first term in the $H_{\rm RKKY}$ describes the indirect exchange interaction between spins mediated by electrons in the $\Gamma$ valley.
Since in the parabolic approximation the electrons at $\Gamma$ valley are free, the first term recovers the asymptotic form $k_Fr\gg 1$ of a classical result for the two-dimensional electron gas.~\cite{Fischer1975, Beal-Monod1987, Litvinov1998}
The second term contains the mixed contribution to the spin susceptibility from both $K$ and $\Gamma$ electrons.
The third term represents the contribution from $K$ valley alone and has the same type of spin-spin interactions as discussed for \ce{MoS2}.~\cite{Parhizgar2013}

\begin{figure}[t]
	\includegraphics[width=\textwidth]{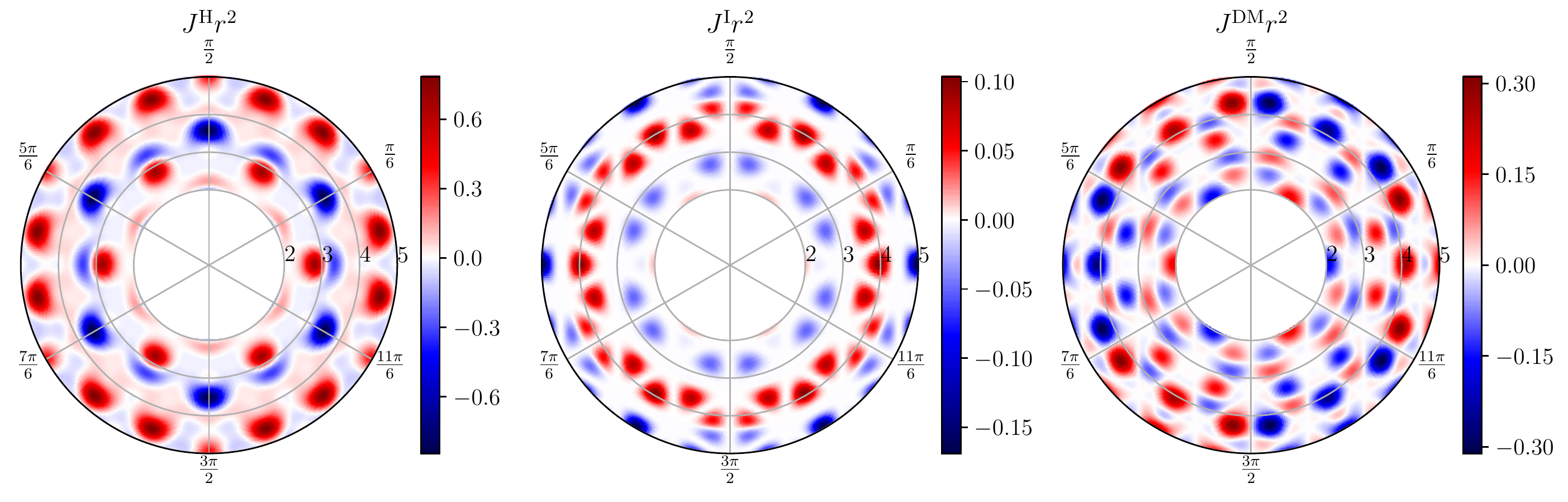}
	\caption{The amplitude of Heisenberg $J^{\rm H}$, Ising $J^{\rm I}$, and Dzyaloshinskii-–Moriya $J^{\rm DM}$ spin-spin interaction for a pair of spins located on the surface of \ce{NbSe2} at distance $r$, with one spin at the origin. 
		The amplitude of $J^X r^2$ is measured in units of $J^2/\Omega_0 \si{eV}$ with primitive unit cell area $\Omega_0=\sqrt 3 a^2/2$. 
		The radial coordinate is in units of lattice constant $a$, and the labeled angles denote the conserved in-plane reflection symmetry axes.
		Since our work is in the limit $r\gg k_F^{-1}\simeq 0.6a$, results for $r<2a$ are excluded from the figure.
	}
	\label{fig:spinspin}
\end{figure}

The Hamiltonian reads after grouping the terms by the spin-spin interactions and summing over $\lambda$ and $\eta$:
\begin{eqnarray}\label{hh_rkky}
H_{\rm RKKY} &=& -\frac{1}{2}\sum_{jl}J^{\rm H}_{jl}\bm S_j\cdot \bm S_l+J^{\rm I}_{jl} S_j^zS_l^z+J^{\rm DM}_{jl}(\bm S_j\times \bm S_l)_z
,\\
J^{\rm H}_{jl} &=& \frac{J^2}{r_{jl}^2}\bigg\{\frac{\rho_{\Gamma(0)}}{2\pi}\sin(2k_F^\Gamma r_{jl})
+\frac{4\rho_\Gamma(0)\rho_K(0)}{3\sqrt{k_F^K k_F^\Gamma}}
\frac{\hbar v_F^K v_F^\Gamma}{v_F^K+v_F^\Gamma}
\sum_n
\cos(\bm K_n\cdot \bm r_{jl})\sin((k_F^\Gamma+k_F^K)r_{jl})\cos(k_{so} r_{jl})\notag\\
&&+\frac{\rho_K(0)}{9\pi}\sum_{nn'}\big[
\cos((\bm K_{n}-\bm K_{n'})\cdot\bm r_{jl})+\cos((\bm K_{n}+\bm K_{n'})\cdot\bm r_{jl})\cos(2k_{so}r_{jl})
\big]\sin(2k_F^Kr_{jl})\bigg\},\notag\\
J^{\rm I}_{ij}&=&-\frac{J^2}{r_{jl}^2}\frac{4\rho_K(0)}{9\pi}\sum_{nn'}\sin(\bm K_n\cdot \bm r_{jl})\sin(\bm K_{n'}\cdot\bm r_{jl})\sin(2k_F^K r_{jl})
\sin^2(k_{so}r_{jl}),\notag\\
J^{\rm DM}_{ij}&=&-\frac{J^2}{r_{jl}^2}\bigg\{
\frac{4\rho_\Gamma(0)\rho_K(0)}{3\sqrt{k_F^K k_F^\Gamma}}
\frac{\hbar v_F^K v_F^\Gamma}{v_F^K+v_F^\Gamma}
\sum_n\sin(\bm K_n\cdot \bm r_{jl})\cos((k_F^\Gamma+k_F^K)r_{jl})\sin(k_{so} r_{jl})
\notag\\
&&+\frac{\rho_K(0)}{9\pi}\sum_{nn'}
\sin((\bm K_n+\bm K_{n'})\cdot\bm r_{jl})\cos(2k_F^Kr_{jl})\sin(2k_{so}r_{jl})
\bigg\}.
\notag
\end{eqnarray}
The final Hamiltonian contain three types of spin-spin interactions: Heisenberg $\bm S_j\cdot \bm S_l$, Ising $ S_j^zS_l^z$, and Dzyaloshinskii-–Moriya (DM) $(\bm S_j\times \bm S_l)_z$ terms.
Some conclusions may be drawn immediately from the classical Hamiltonian~\eqref{hh_rkky}.
The first two terms favor collinear ordering of spins, while the DM term favors orders where spins rotate in the $xy$ plane.
The classical Hamiltonian also shows that the spin-spin interactions decay with square distance between the spins, as in the two-dimensional electron gas, but, in contrast, the interactions are anisotropic in space due to the $K$ valley electrons from \ce{NbSe2}, similar to \ce{MoS2}.~\cite{Parhizgar2013}
The oscillations are roughly with a period $r=\pi/k_F$.
As seen in Fig.~\ref{fig:spinspin}, the interactions obey the underlying $C_3$ symmetry in \ce{NbSe2}.

The DM and Ising terms exist only in presence of Ising spin-orbit coupling.
Only the spin-rotational invariant Heisenberg interactions survive, when $k_{so}\to0$ in Eq.~\eqref{hh_rkky}.
Due to vanishing spin-orbit splitting on the conserved in-plane reflection symmetry axes (see App.~\ref{app:tb}), the DM and Ising terms also vanish along them (see Fig.~\ref{fig:spinspin}).

In general the three terms in $H_{\rm RKKY}$ compete to set the ground state's ordering of spins.
Let us start with the case of two classical spins to get a sense of the competing orders in our system.
The numerical results are displayed in Fig.~\ref{fig:spinspin} showing that the amplitudes of the three terms are on the same order of magnitude (especially going to even larger distances that shown here).
It is immediately apparent that interactions obey the $C_3$ symmetry of \ce{NbSe2} and that Ising and DM interactions vanish on the conserved in-plane reflection symmetry axes.
For a particular choice of distance between spins and their position on the \ce{NbSe2} surface, there are situations where DM term dominates, so in those cases the ansatz in Eq.~\eqref{h1} would not be correct. 
It is interesting to note that for the case $r=3a$ and $\phi=0$, which is the focus of our numerical study in Sec.~\ref{sec:num_res}, the DM and Ising interactions almost vanish, leaving a ferromagnetic Heisenberg interaction $J^{\rm H}>0$ in accordance to the ansatz of Eq.~\eqref{h1}.


An exhaustive study of all possible spin orderings is outside the scope of the present paper.
As we have seen the RKKY interaction is direction dependent, so the ground state spin orientation will depend both on the distance between spins and the chain orientation.
In the following, we will minimize the RKKY Hamiltonian starting from a classical ansatz for spin ordering: a transverse conical magnetic structure which allows to investigate the competition between the Heisenberg and the DM interactions.
Note that cycloidal, with spins rotating in a plane normal to the \ce{NbSe2} plane, or helical ordering, with spins rotating in a plane normal to the impurity chain axis, will result in vanishing DM interactions, since $(\bm S_j\times \bm S_l)_z=0$ in both cases.

\begin{figure}[t]
	\includegraphics[width=0.4\textwidth]{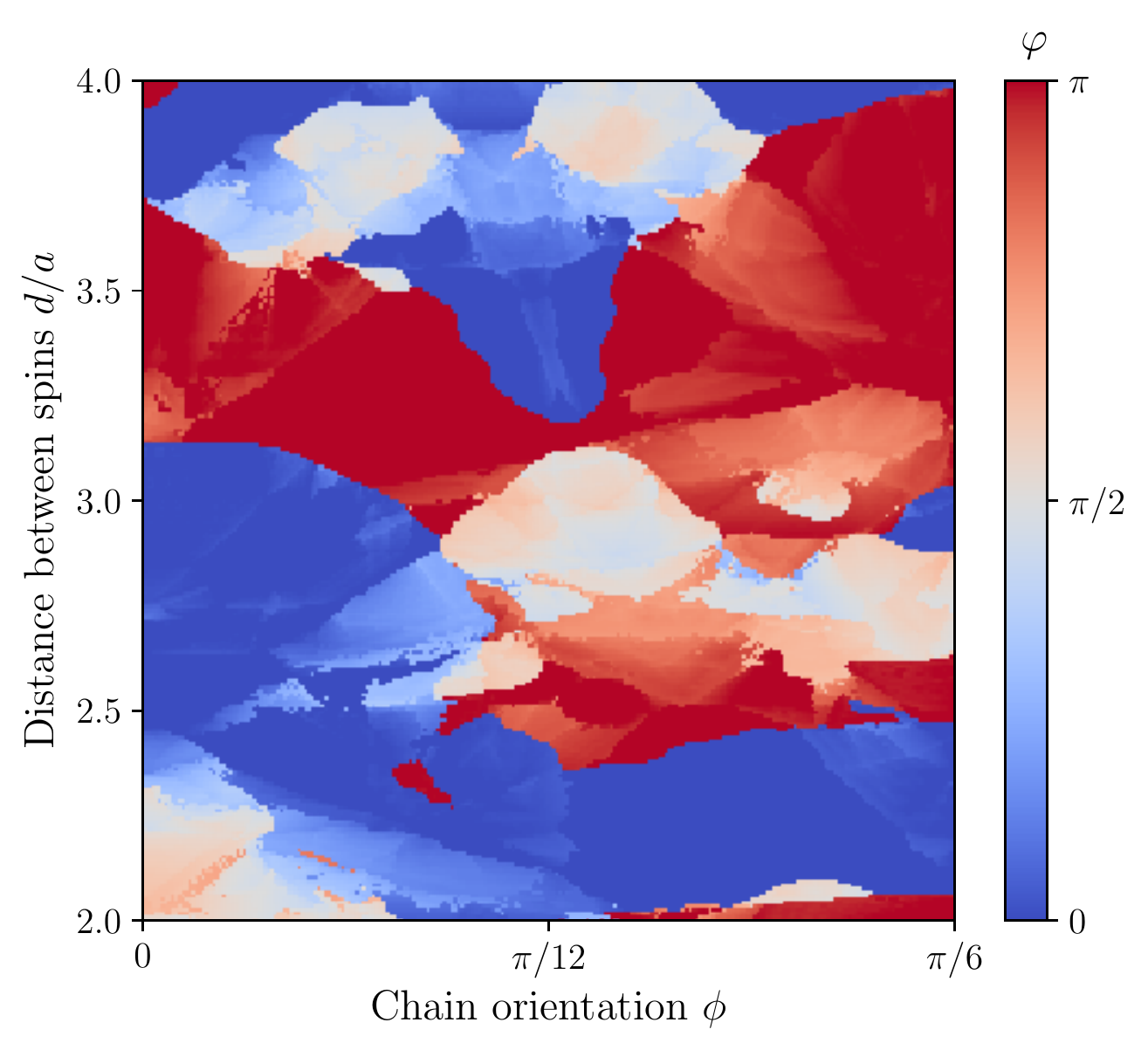}
	\caption{Classical ground state for a chain of spins on \ce{NbSe2} surface as a function of the angle $\phi$ (between the spin chain and the $x$ axis) and the distance $d$ between spins in units of lattice constant $a$. 
		Ferromagnetic and antiferromagnetic order are realized for $\varphi=0$, respectively, $\varphi=\pi$, with in-plane helical order for other values.}
	\label{fig:class_gs}
\end{figure}

The spins precess around the $\hat{\bm z}$ axis with a polar angle $\varphi$, at a fixed azimuthal angle $\theta$.
The transverse conical structure is generated by spins:
\begin{equation}
\bm S_j/S = \cos\theta \hat{\bm z}+\sin\theta \cos(j\varphi)\hat{\bm r}
+\sin\theta \sin(j\varphi)(\hat{\bm z}\times\hat{\bm r}),
\end{equation}
where $\hat{\bm r}$ points in $xy$ plane along the impurity chain.
Initial numerical study shows in this case a preference for $\theta=\pi/2$ orientation.
Therefore we simplify the problem by fixing $\theta=\pi/2$ and finding the minimum energy under a variation of $\varphi$ in the interval $[0, \pi]$.
Note that ferro- and antiferromagnetic order is realized for $\varphi=0$, and, respectively, $\varphi=\pi$, with in-plane $xy$ helical ordering for intermediate values.
The RKKY Hamiltonian becomes in this case, for $N\to \infty$ impurities:
\begin{equation}
H_{\rm RKKY} = -NS^2\sum_{j=1}^\infty \big[
J^{\rm H}_{j0}\cos(j\varphi)
-J^{\rm DM}_{j0} \sin(j\varphi)
\big].
\end{equation}
The sum is over distances $jd$ between the spins, with terms decaying as $j^{-2}$.
Therefore in the minimization process it is sufficient to truncate the sum to a finite number of terms (here 50).
The results are shown in Fig.~\ref{fig:class_gs}.
There are large areas in parameter space where collinear order is preferred.
Note that, in contrast, a two-dimensional metal with Rashba spin-orbit coupling does not stabilize collinear order of spin.
Instead additional crystal field effects, which at first order favor an Ising coupling of spins, are invoked to produce a collinear order.~\cite{Heimes2015}
More dedicated numerical studies are required to include these effects, to decide whether they are enough to stabilize ferromagnetic order for any chain orientation, and whether the classical ground state is stable to quantum and thermal fluctuations~\cite{Heimes2015, Kim2014}.

\section{Effect of superconducting coherence length on the topological phase diagram}
\label{app:topo}
\begin{figure}[t]
\includegraphics[width=\columnwidth]{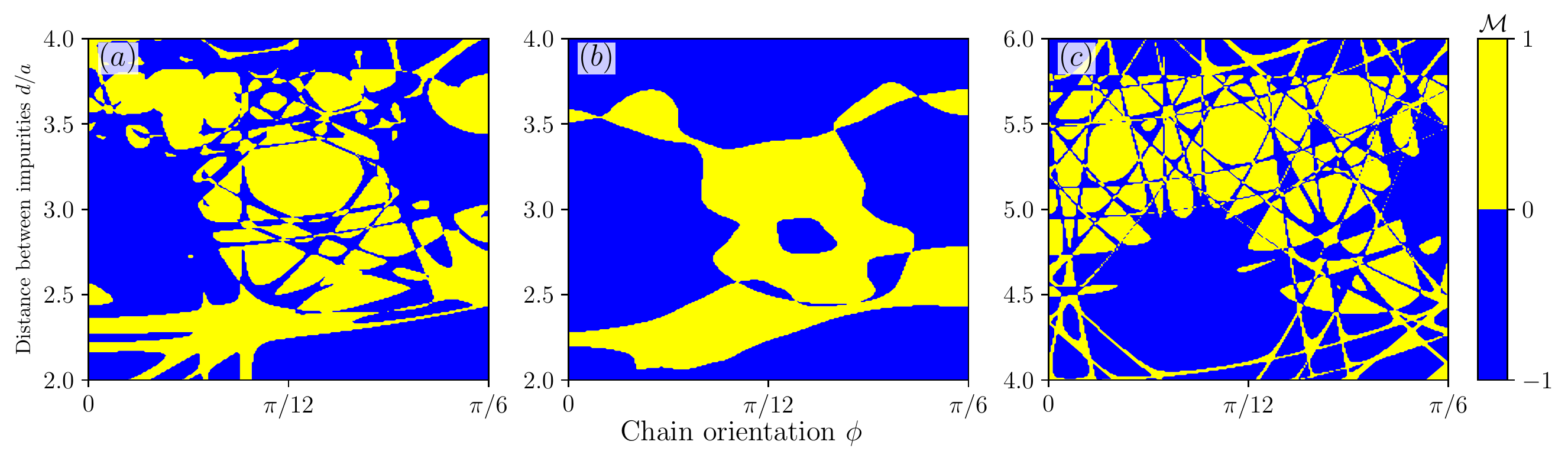}
\caption{[(a) and (b)] Topological phase diagram when increasing in factors of ten the superconducting gap (decreasing the coherence length) with respect with the physical $\Delta=\SI{1}{meV}$ from Fig.~\ref{fig:topdiag}.
(a) $\Delta = \SI{10}{meV}$, used also in the numerical study of Sec.~\ref{sec:num_res}.
(b) $\Delta = \SI{0.1}{eV}$.
(c) Topological phase diagram at physical $\Delta=\SI{1}{meV}$, but at larger distances between impurities, $d$ in $[4,6]a$. This could prove relevant for YSR generated by magnetic molecules such as phthalocyanines and porphyrins, which, when functionalized, have linear lengths comparable to $5a$.
		Magnetic exchange energy is fixed at $JS=2/\pi\rho(0)$.}
	\label{fig:top_diag_diff}
\end{figure}

This appendix investigates the behavior of the phase diagram as the coherence length is gradually decreased.
The coherence length controls the number of distant-neighbor hopping terms relevant in the Hamiltonian.
Reducing the coherence length, or equivalently increasing $\Delta$, simplifies the phase diagram, which is seen to gradually losing its fine structure as Hamiltonian matrix elements are less oscillating in momentum space [Figs.~\ref{fig:topdiag}, \ref{fig:top_diag_diff}(a) and \ref{fig:top_diag_diff}(b), where $\Delta$ is, respectively, \SI{1}{meV}, \SI{10}{meV}, and \SI{100}{meV}].
The topological phases tend to grow at smaller coherence length. A particular case is in the unphysical short coherence length limit $\xi_0\ll d$, such that the effective Hamiltonian has only nearest-neighbor hoppings.
The Bloch Hamiltonian follows directly from Eq.~\eqref{bloch_eff} under the approximation of polylogarithmic functions with their argument.
This leads at $\alpha=1$ or $JS=2/\pi\rho(0)$ to a topological phase extension to the entire allowed [where $h(k)$ is not vanishing] parameter space in $d$ and $\phi$, since $h(k)$ changes sign between $k=0$ and $k=\pi/d$ due to a $\cos(kd)$ behavior:
\begin{equation}
\frac{h(k)}{\Delta} = -2\cos(k d)\bigg[\frac{\rho_\Gamma(0)}{\rho(0)}\sqrt{\frac{2}{\pi k_F^\Gamma d}}
\cos\big(k_F^\Gamma d-\frac{\pi}{4}\big)e^{-\frac{d}{\xi_0^\Gamma}}
+\frac{1}{3}
\frac{\rho_{\eta\lambda}(0)}{\rho(0)}
\sqrt{\frac{2}{\pi k_F^K d}}\sum_{n\eta\lambda}
\cos\big(k_F^K d-\frac{\pi}{4}\big)e^{-\frac{d}{\xi_0^K}}
\bigg].
\end{equation}

Additionally, Fig.~\ref{fig:top_diag_diff}(c) shows the topological phase diagram for larger distances between magnetic impurities, on the order expected for molecular chains of magnetic porphyrins and phthalocyanines.

\twocolumngrid
\bibliography{bibl}

\end{document}